\documentclass{lmcs} 
\pdfoutput=1

\usepackage{lastpage}
\lmcsdoi{15}{3}{33}
\lmcsheading{}{\pageref{LastPage}}{}{}%
{Jul.~03,~2018}{Sep.~30,~2019}{}


\usepackage{hyperref}

\usepackage{dsfont}
\usepackage{amsmath,amssymb}
\usepackage{stmaryrd}
\usepackage{fancyhdr}
\usepackage{cite}
\usepackage{tabularx}
\usepackage{environ}
\usepackage{lipsum}
\usepackage{xspace}
\usepackage{textcomp}
\usepackage[pdflatex,recompilepics=false]{gastex}
\usepackage{booktabs}

\usepackage{ifpdf}

\ifpdf
\usepackage{tikz}
\usetikzlibrary{arrows,automata}
\usetikzlibrary{arrows, automata, positioning,calc}
\usetikzlibrary{shapes}
\fi

\makeatletter
\renewcommand{\paragraph}{\@startsection{paragraph}{6}{\z@}{2ex}{-0.7em}{\normalsize\bf}}
\makeatother

\makeatletter
\newcommand{\problemtitle}[1]{\gdef\@problemtitle{#1}}
\newcommand{\probleminput}[1]{\gdef\@probleminput{#1}}
\newcommand{\problemquestion}[1]{\gdef\@problemquestion{#1}}
\NewEnviron{decproblem}{
  \problemtitle{}\probleminput{}\problemquestion{}
  \BODY
  \par\addvspace{0.8\baselineskip}
  \noindent
  \fbox{\centering
  \begin{tabular}{rl}
    \multicolumn{2}{l}{\normalsize\@problemtitle} \\
    \hline 
    \rule{0pt}{3ex}
    \normalsize\textbf{Input:} & \normalsize\@probleminput\\[0.5ex]
    \normalsize\textbf{Question:} & \normalsize\@problemquestion
  \end{tabular}}
  \par\addvspace{0.8\baselineskip}
}

\newcommand{\ie}{i.e.}
\newcommand{\etal}{et al.}

\newcommand{\nexptime}{\mathsf{NEXPTIME}}

\newcommand{\freezeLTL}{freeze~LTL\xspace}
\newcommand{\FreezeLTL}{Freeze~LTL\xspace}

\newcommand{\modelsex}{\models_\exists}

\newcommand{\Props}{\mathbb{P}}
\renewcommand{\prop}{p}

\newcommand{\propmap}{\mu}

\newcommand{\adel}{\Box}

\newcommand{\bound}{d}
\newcommand{\ptest}[2]{\mathord{#1\,}#2}
\newcommand{\ctest}[2]{\mathord{#1\,}#2}
\newcommand{\bowtietest}[1]{\mathord{\bowtie\,}#1}
\newcommand{\qinit}{q_\mathsf{in}}

\newcommand{\btrue}{\texttt{true}}
\newcommand{\bfalse}{\texttt{false}}
\newcommand{\afatrans}[3]{#1 \xrightarrow{#2} #3}
\newcommand{\seen}[1]{\checkmark_{\!\!#1}}

\newcommand{\FLTL}{\ensuremath{\textup{LTL}^\downarrow}}
\newcommand{\FFLTL}{\ensuremath{\flat\textup{LTL}^\downarrow}}

\newcommand{\coFFLTL}{\ensuremath{\textup{co-}\FFLTL}}

\newcommand{\LTL}{\textup{LTL}}

\newcommand{\Reg}{\mathcal{R}}
\newcommand{\dfreeze}[2]{\mathord{\downarrow_{#1}} #2}

\newcommand{\regtest}[2]{\mathord{#1\,}{#2}}

\newcommand{\neXt}[1]{\mathsf{X} #1}
\newcommand{\Until}[2]{#1 \,{\mathsf{U}}\, #2}
\newcommand{\Untilwp}{{\mathsf{U}}}
\newcommand{\Releasewp}{{\mathsf{R}}}
\newcommand{\Globally}{\mathsf{G}}
\newcommand{\Future}{\mathsf{F}}
\newcommand{\Release}[2]{#1 \mathrel{\mathsf{R}} #2}

\newcommand{\Class}{\mathcal{C}}
\newcommand{\Logic}{\mathcal{L}}

\newcommand{\Reachpred}[2]{\ensuremath{#1 \models \mathit{Reach}(#2)}}
\newcommand{\RepReachpred}[2]{\ensuremath{#1 \models \mathit{Reach}^\omega(#2)}}
\newcommand{\Tests}{\textup{Tests}}

\newcommand{\Conf}[1]{\mathcal{C}_{#1}}

\newcommand{\Nonemptiness}[1]{\ensuremath{#1\textsc{-B{\"u}chi}}}
\newcommand{\Reach}[1]{\ensuremath{#1\textsc{-Reachability}}}
\newcommand{\Buechi}[1]{#1-\textsc{B{\"u}chi}}
\newcommand{\MC}[2]{\ensuremath{#1\textsc{-MC}(#2)}}

\newcommand{\bcsep}[1]{\#_{#1}}
\newcommand{\transl}[1]{[#1]}
\newcommand{\zerobit}{\mathbf{0}}
\newcommand{\onebit}{\mathbf{1}}
\newcommand{\firstsep}{\mathit{first}}
\newcommand{\lastsep}{last}
\newcommand{\eqsuff}[1]{\mathit{suff}_{\!#1}^=}
\newcommand{\jump}[2]{\mathord{\curvearrowright_{#1}}{#2}}
\newcommand{\newp}{s}
\newcommand{\news}{\hat{s}}
\newcommand{\newR}{T}
\newcommand{\newr}{t}
\newcommand{\newxp}{y}

\newcommand{\search}[1]{\mathsf{search}(#1)} 
\newcommand{\searchn}[1]{\mathsf{search}^+(#1)} 
\newcommand{\present}[1]{\mathsf{present}(#1)} 
\newcommand{\gotoright}[1]{\mathsf{right}(#1)} 
\newcommand{\gotoleft}[1]{\mathsf{left}(#1)} 

\newcommand{\regassign}{\nu}

\newcommand{\height}{\parbox[0pt][5.5ex][c]{0cm}{}}
\newcolumntype{L}[1]{>{\height\centering\arraybackslash}l{#1}}
\newcolumntype{C}[1]{>{\height\centering\arraybackslash}c{#1}}
\newcolumntype{R}[1]{>{\height\centering\arraybackslash}r{#1}}
\newcolumntype{P}[1]{>{\centering\arraybackslash}p{#1}}

\renewcommand{\epsilon}{\varepsilon}
\renewcommand{\phi}{\varphi}

\newcommand{\B}{\mathcal{B}}
\newcommand{\N}{\mathds{N}}
\newcommand{\Z}{\mathbb{Z}}

\newcommand{\succinct}{\textup{\textsf{S}}}
\newcommand{\ptests}{\textup{\textsf{P}}}
\newcommand{\ctests}{\textup{\textsf{C}}}

\newcommand{\socapc}{\textup{OCA(\succinct,\ptests,\ctests)}\xspace} 
\newcommand{\socap}{\textup{OCA(\succinct,\ptests)}\xspace} 
\newcommand{\ocapc}{\textup{OCA(\ptests,\ctests)}\xspace} 
\newcommand{\ocap}{\textup{OCA(\ptests)}\xspace} 
\newcommand{\soca}{\textup{OCA(\succinct)}\xspace} 
\newcommand{\oca}{\textup{OCA}\xspace}

\newcommand{\ocaps}{\ocap}

\newcommand{\ocas}{\text{OCA}\xspace}

\newcommand{\POCA}{\mathcal{A}}

\newcommand{\df}{:=}

\newcommand{\zero}{\ctest{=}{0}}

\newcommand{\first}{\mathsf{first?}}
\newcommand{\true}{\mathsf{true}}

\newcommand{\inst}{\gamma}

\newcommand{\pspace}{\mathsf{PSPACE}}

\makeatletter
\newcommand*{\defeq}{\mathrel{\rlap{%
                     \raisebox{0.3ex}{$\m@th\cdot$}}%
                     \raisebox{-0.3ex}{$\m@th\cdot$}}%
                     =}
\makeatother

\newcommand{\RP}{\ensuremath{\mathbb{R}_{\ge 0}}}

\newcommand{\loc}{s}
\newcommand{\locs}{S}

\newcommand{\clocks}{\mathcal{C}}
\newcommand{\clock}{\mathfrak{c}}

\newcommand{\OCA}{\mathcal{M}}

\newcommand{\op}{\mathsf{op}}

\newcommand{\params}{\mathcal{X}}
\newcommand{\paramwords}[1]{\mathcal{W}_{#1}}
\newcommand{\Params}{\mathcal{X}}
\newcommand{\param}{x}
\newcommand{\const}{c}
\newcommand{\NP}{\mathsf{NP}}
\newcommand{\NEXPTIME}{\mathsf{NEXPTIME}}

\newcommand{\tNEXPTIME}{\mathsf{2NEXPTIME}}
\newcommand{\NLOGSPACE}{\mathsf{NLOGSPACE}}
\newcommand{\EXPSPACE}{\mathsf{EXPSPACE}}
\newcommand{\PSPACE}{\pspace}
\newcommand{\pocatrans}[1]{\longrightarrow_{#1}}
\newcommand{\pocatransp}[1]{\longrightarrow_{#1}}
\newcommand{\ocatrans}{\longrightarrow}

\newcommand{\twoafa}{\mathcal{T}}

\newcommand{\UMC}[2]{\ensuremath{#1\textsc{-MC}^\forall(#2)}}

\newcommand{\circuit}{\mathbb{C}}

\def\cf{\emph{cf.}}

\begin{document}

\title[The Complexity of Flat Freeze LTL]{The Complexity of Flat Freeze LTL\rsuper{*}}

\titlecomment{{\lsuper*}This work has been partly supported by the  ANR  research  program PACS (ANR--14--CE28--0002) and by DFG, project QU 316/1--2. A preliminary version of this paper appeared as~\cite{bollig-complexity-17}.}

\author[B.~Bollig]{Benedikt Bollig\rsuper{a}}
\address{\lsuper{a}CNRS, LSV, ENS Paris-Saclay, Universit{\'e} Paris-Saclay}
\email{bollig@lsv.fr}

\author[K.~Quaas]{Karin Quaas\rsuper{b}}
\address{\lsuper{b}Universit\"at Leipzig}
\email{quaas@informatik.uni-leipzig.de}

\author[A.~Sangnier]{Arnaud Sangnier\rsuper{c}}
\address{\lsuper{c}IRIF, Universit{\'e} Paris Diderot}
\email{sangnier@irif.fr}





\begin{abstract}
We consider the model-checking problem for \freezeLTL on one-counter automata (\ocas). Freeze LTL extends LTL with the freeze quantifier, which allows one to store different counter values of a run in registers so that they can be compared with one another. As the model-checking problem is undecidable in general, we focus on the flat fragment of \freezeLTL, in which the usage of the freeze quantifier is restricted. In a previous work, Lechner et al.\ showed that model checking for flat \freezeLTL on \ocas with binary encoding of counter updates is decidable and in $\tNEXPTIME$. In this paper, we prove that the problem is, in fact, $\NEXPTIME$-complete no matter whether counter updates are encoded in unary or binary. Like Lechner et al., we rely on a reduction to the reachability problem in \ocas with parameterized tests (\ocaps). The new aspect is that we simulate \ocaps by alternating two-way automata over words. This implies an exponential upper bound on the parameter values that we exploit towards an $\NP$ algorithm for reachability in \ocaps with unary updates. We obtain our main result as a corollary.
As another application, relying on a reduction by Bundala and Ouaknine, one obtains an alternative proof of the known fact that reachability in closed parametric timed automata with one parametric clock is in $\NEXPTIME$.
\end{abstract}

\maketitle

\section{Introduction}

One-counter automata (\ocas) are a simple yet fundamental computational model operating on a single counter that ranges over the non-negative integers.
Albeit being a classical model, \ocas are in the focus of ongoing research in the verification community (\cf, for example,~\cite{HaaseKOW09,GollerHOW12,LechnerMOPW16,DemriG09,GollerHOW10,GollerL13,HofmanLMT16}).
A large body of work is devoted to model checking \ocas, \ie, the question whether all runs of a given \oca satisfy a temporal-logic specification.
A natural way to model runs of \ocas is in terms of \emph{data words}, \ie, words where every position carries two pieces of information: a set of propositions from a finite alphabet, and a datum from an infinite alphabet. In our case, the datum represents the current counter value. Now, reasoning about the sequence of propositions that is produced by a run amounts to model checking \ocas against classical temporal logics like linear-time temporal logic (LTL) and computation-tree logic (CTL)~\cite{GollerL13,GollerHOW12,GollerHOW10,Haase12}.
But it is natural to also reason about the unboundedly many counter values that may occur.
Several formalisms have been introduced that can handle infinite alphabets, including variants or extensions of monadic second-order logic, LTL, and CTL~\cite{DemriL09,DemriLS10,BojanczykDMSS11,FengLQ15}.
\FreezeLTL is an extension of LTL that allows one to remember, in terms of the \emph{freeze quantifier}, certain counter values for later comparison in a run of the \oca under consideration (\cf~\cite{French03,LisitsaP05,DemriLN07,DemriL09}).
Unfortunately, satisfiability and model checking \ocas against \freezeLTL are undecidable~\cite{DemriL09,DemriLS10}.

In this paper, we study model checking of \ocas against formulas in \emph{flat \freezeLTL}, a fragment of \freezeLTL that restricts the freeze quantifier, but allows for unlimited usage of comparisons.
A typical property definable in flat \freezeLTL is ``there exists a counter value that occurs infinitely often''.
Moreover, the negation of many natural \freezeLTL specifications can be expressed in flat \freezeLTL\@.
The approach of restricting the syntax of a temporal logic to its flat fragment has also been pursued in~\cite{ComonC00} for Constraint LTL, and in~\cite{BouyerMOW08} for MTL\@.

Demri and Sangnier~\cite{DS-fossacs10} reduced model checking \ocas against flat \freezeLTL to reachability in \ocas \emph{with parameterized tests} (\ocaps).
In an \ocap, counter values may be compared with parameters whose values are arbitrary but fixed. Reachability asks whether a given state can be reached under \emph{some} parameter instantiation. Essentially, the translation of a flat \freezeLTL formula into an \ocap interprets every freeze quantifier as a parameter, whose value can be compared with counter values arbitrarily often. Decidability of the reachability problem for \ocaps, however, was left open. Recently, Lechner \etal\ proved decidability by a reduction to satisfiability in Presburger arithmetic~\cite{LechnerMOPW16}. As a corollary, they obtain a $\tNEXPTIME$ upper bound for model checking \ocas against formulas in flat \freezeLTL, assuming that counter updates in \ocas are encoded in binary.

Our main technical contribution is an improvement of the result by Lechner \etal~We proceed in two main steps.
First, we show that reachability for \ocaps (with unary counter updates) can be reduced to non-emptiness of alternating two-way (finite) automata.
Interestingly, alternating two-way automata have already been used as an algorithmic framework for game-based versions of pushdown processes~\cite{KupfermanV00,Cachat01}, one-counter processes~\cite{Serre06}, and timed systems~\cite{ABGMN-fi13}.
Our link already implies decidability of both reachability for \ocaps (in $\PSPACE$ when counter updates are unary) and model checking \ocas against flat \freezeLTL (in $\EXPSPACE$). However, we can go further.
First, we deduce an exponential upper bound on the largest parameter value needed for an accepting run in the given \ocap. Exploiting this bound and a technique by Galil~\cite{Galil76}, we show that, actually, reachability for \ocaps is in $\NP$.
As a corollary, we obtain a $\NEXPTIME$ upper bound for model checking \ocas against flat \freezeLTL\@. Using a result from~\cite{GollerHOW10}, we can also show $\NEXPTIME$-hardness, which establishes the precise complexity of the model-checking problem.
Our result applies no matter whether counter updates in the given \oca are encoded in unary or binary.

\paragraph{Outline.} In Section~\ref{sec:prel}, we define \ocas, (flat) \freezeLTL, and the corresponding model-checking problems. Section~\ref{sec:poca-np} is devoted to reachability in \ocaps, which is at the heart of our model-checking procedures. The reduction of model checking to reachability in \ocaps is given in Section~\ref{sec:soca-mc}, where we also present lower bounds.
In Section~\ref{sec:soca-universal-mc}, we consider the \emph{universal} model-checking problem. Finally, in Section~\ref{sec:pta}, we apply our results to show that reachability in parametric timed automata with one parametric clock (1PTA) and closed guards (which corresponds to the discrete-time case) is in $\NEXPTIME$, which matches the known lower bound. Note that this upper bound had already been shown in~\cite{BenesBLS15} for general 1PTA  (over both discrete time and continuous time), using a different proof technique. We conclude in Section~\ref{sec:conclusion}.

\section{Preliminaries}\label{sec:prel}

\newcommand{\Trans}{\Delta}
\newcommand{\updTrans}{\Delta_{\mathsf{upd}}}
\newcommand{\pTrans}{\Delta_{\mathsf{param}}}
\newcommand{\cTrans}{\Delta_{\mathsf{const}}}

\subsection{One-Counter Automata with Parameterized Tests}

We start by defining one-counter automata \emph{with all extras} such as succinct encodings of updates, parameters, and comparisons with constants. Well-known subclasses are then identified as special cases.

For the rest of this paper, we fix a countably infinite set $\Props$ of \emph{propositions}, which will label states of a one-counter automaton. Transitions of an automaton may perform tests to compare the current counter value with zero, with a parameter, or with a constant.

\begin{defi}
A \emph{one-counter automaton with succinct updates, parameterized tests, and comparisons with constants}, \socapc for short, is a tuple $\POCA = (Q,\params,\qinit,\Delta,\propmap)$ where
\begin{itemize}\itemsep=0.5ex
\item $Q$ is a finite set of \emph{states},
\item $\params$ is a finite set of \emph{parameters} ranging over $\N$,
\item $\qinit \in Q$ is the \emph{initial state},
\item $\Trans = \updTrans \uplus \pTrans \uplus \cTrans$ is the finite set of \emph{transitions}, which is partitioned into
\begin{itemize}
\item $\updTrans \subseteq Q \times \Z \times Q$,
\item $\pTrans \subseteq Q \times \{\ptest{<}{\param}\,, \ptest{=}{\param}\,,\ptest{>}{\param} \mid \param\in\params\} \times Q$,
\item $\cTrans \subseteq Q \times \{\ctest{<}{\const}\,, \ctest{=}{\const}\,,\ctest{>}{\const} \mid \const\in\N\} \times Q$, and
\end{itemize}
\item $\propmap: Q \to 2^\Props$ maps each state to a \emph{finite} set of propositions.
\end{itemize}
We define the size of $\POCA$ as
\[|\POCA| =
\left(
\begin{array}{rl}
& |Q| + |\params| + |\Trans| + \sum_{q \in Q} |\propmap(q)|\\[1ex]
+ & \sum \{\log(|z|) \mid (q,z,q') \in \updTrans \text{ with } z\neq 0\bigr\}\bigr\}\\[1ex]
+ & \sum \{\log(\const) \mid (q,\bowtietest{\const},q') \in \cTrans \text{ with } \const > 0\bigr\}
\end{array}\right)\,.
\]
\end{defi}

Let $\Conf{\POCA} \df Q \times \N$ be the set of \emph{configurations} of $\POCA$. In a configuration $(q,v) \in \Conf{\POCA}$, the first component $q$ is the current state and $v$ is the current counter value, which is always non-negative. The semantics of $\POCA$ is given w.r.t.\ a \emph{parameter instantiation} $\inst: \params \to \N$ in terms of a global transition relation ${\pocatrans{\inst}} \subseteq \Conf{\POCA} \times \Conf{\POCA}$. For two configurations $(q,v)$ and $(q',v')$, we have $(q,v) \pocatransp{\inst}  (q',v')$ if there is $(q,\mathsf{op},q') \in \Delta$ such that one of the following holds:
\begin{itemize}
\item $\op\in\Z$ and $v' = v + \op$,

\item $\op = \bowtietest{\param}$ and $v = v'$ and $v \bowtie \inst(\param)$, for some $\param \in \Params$ and ${\bowtie} \in\{<,=,>\}$, or

\item $\op = \bowtietest{\const}$ and $v = v'$ and $v \bowtie \const$, for some $\const \in \N$ and ${\bowtie} \in\{<,=,>\}$.
\end{itemize}

\noindent
Thus, transitions of an \socapc either increment/decrement the counter by a value $z \in \Z$, or compare the counter value with a parameter value or a constant. Note that, in this model, the counter values are always positive. Moreover, we assume a binary encoding of constants and updates.

A $\inst$-\emph{run} of $\POCA$ is a finite or infinite sequence $\rho = (q_0,v_0) \pocatrans{\inst}  (q_1,v_1)  \pocatrans{\inst} \cdots $ of global transitions (a run may consist of one single configuration $(q_0,v_0)$). We say that $\rho$ is \emph{initialized} if $q_0=\qinit$ and $v_0=0$.

\medskip

We identify some well-known special cases of the general model. Let $\POCA=(Q,\params,\qinit,\Delta,\propmap)$ be an \socapc. We say that
\begin{itemize}\itemsep=0.5ex
\item $\POCA$ is an \ocapc if $\updTrans \subseteq Q \times \{+1,0,-1\} \times Q$ (i.e., counter updates are \emph{unary});

\item $\POCA$ is an \socap if $\cTrans \subseteq Q \times \{\zero\} \times Q$ (i.e., only comparisons with $0$ are allowed);

\item $\POCA$ is an \ocap if $\updTrans \subseteq Q \times \{+1,0,-1\} \times Q$ and $\cTrans \subseteq Q \times \{\zero\} \times Q$ (i.e., counter updates are unary and only comparisons with $0$ are allowed);

\item $\POCA$ is an \soca if $\params = \emptyset$ (i.e., $\pTrans = \emptyset$) and
$\cTrans \subseteq Q \times \{\zero\} \times Q$\,(i.e., there is no parameter and only comparisons with $0$ are allowed);

\item $\POCA$ is an \oca if $\params = \emptyset$ and, moreover, all transition labels are among $\{+1,0,-1\} \cup \{\zero\}$ (i.e., there is no parameter, counter updates are \emph{unary} and only comparisons with $0$ are allowed).
\end{itemize}

\noindent
We may omit $\params$ when it is equal to $\emptyset$ (\ie, in the case of an \soca or \oca)  and simply refer to $(Q,\qinit,\Delta,\propmap)$. Since, in this case, the global transition relation does not depend on a parameter instantiation anymore, we may just write $\pocatrans{}$ instead of $\pocatrans{\inst}$. Similarly, for reachability problems (see below), $\propmap$ is irrelevant so that it can be omitted, too.

\medskip

One of the most fundamental decision problems we can consider is whether a given state $q_f \in Q$ is reachable. Given some parameter instantiation $\inst$, we say that $q_f$ is $\inst$-\emph{reachable} if there is an initialized run
$(q_0,v_0) \pocatrans{\inst} \cdots \pocatrans{\inst}  (q_n,v_n)$ such that $q_n = q_f$.
Moreover, $q_f$ is \emph{reachable}, written $\Reachpred{\POCA}{q_f}$, if it is $\inst$-reachable for some $\inst$.
Now, for a class $\Class \in \{\socapc,\ocapc,\socap,\ocap,\soca,\oca\}$, the \emph{reachability problem} is defined as follows:
\begin{center}
\begin{decproblem}
  \problemtitle{\Reach{\Class}}
  \probleminput{$\POCA = (Q,\params,\qinit,\Delta) \in \Class$ and $q_f \in Q$}
  \problemquestion{Do we have $\Reachpred{\POCA}{q_f}$\,?}
\end{decproblem}
\end{center}

Towards the \emph{B{\"u}chi problem} (or \emph{repeated reachability}), we write $\RepReachpred{\POCA}{q_f}$ if there are a parameter instantiation $\inst$ and an infinite initialized $\inst$-run $(q_0,v_0) \pocatrans{\inst}  (q_1,v_1) \pocatrans{\inst} \cdots$ such that $q_i = q_f$ for infinitely many $i \in \N$.
The B{\"u}chi problem asks whether some state from a given set $F \subseteq Q$ can be visited infinitely often:

\begin{center}
\begin{decproblem}
  \problemtitle{\Nonemptiness{\Class}}
  \probleminput{$\POCA = (Q,\params,\qinit,\Delta) \in \Class$ and $F \subseteq Q$}
  \problemquestion{Do we have $\RepReachpred{\POCA}{q_f}$ for some $q_f \in F$\,?}
\end{decproblem}
\end{center}

\noindent
It is known that \Reach{\oca} and \Nonemptiness{\oca} are $\NLOGSPACE$-complete~\cite{Valiant1975,DemriG09}, and that \Reach{\soca} and \Nonemptiness{\soca} are $\NP$-complete~\cite{HaaseKOW09,Haase12} (cf.\ also Table~\ref{table:results}).

\subsection{Freeze LTL and its Flat Fragment}

We now define \freezeLTL\@. To do so, we fix a countably infinite supply of registers $\Reg$.

\begin{defi}[Freeze LTL]
The logic \emph{\freezeLTL}, denoted by $\FLTL$, is given by the grammar
\begin{align*}
\phi &~~ ::= ~~ \prop ~~\mid~~ \regtest{\bowtie\,}{r} ~~\mid~~ \neg\phi ~~\mid~~ \phi \vee \phi ~~\mid~~ \phi \wedge \phi ~~\mid~~ \neXt{\phi} ~~\mid~~ \Until{\phi}{\phi} ~~\mid~~ \Release{\phi}{\phi} ~~\mid~~ \dfreeze{r}{\phi}
\end{align*}
where $\prop \in \Props$, $r \in \Reg$, and $\mathord{\bowtie} \in \{<,=,>\}$.
\end{defi}

Note that, apart from the until operator $\Untilwp$, we also include the dual release operator $\Releasewp$. This is convenient in proofs, as it allows us to assume formulas to be in negation normal form. We also use common abbreviations such as $\phi \to \psi$ for $\neg \phi \vee \psi$, $\Future \phi$ for $\Until{\true}{\phi}$ (with $\true = p \vee \neg p$ for some proposition $p$), and $\Globally \phi$ for $\neg\Future\neg\phi$.

We call $\dfreeze{r}{}$ the \emph{freeze quantifier}. It stores the current counter value in register $r$. Atomic formulas of the form $\regtest{\bowtie\,}{r}$, called \emph{register tests}, perform a comparison of the current counter value with the contents of $r$, provided that they are in the scope of a freeze quantifier. Actually, inequalities of the form $\regtest{<}{r}$ and $\regtest{>}{r}$ were not present in the original definition of \FLTL, but our techniques will take care of them without extra cost or additional technical complication.
Classical $\LTL$ is obtained as the fragment of $\FLTL$ that uses neither the freeze quantifier nor register tests.

A formula $\phi \in \FLTL$ is interpreted over an infinite sequence $w = (P_0,v_0)(P_1,v_1) \ldots \in {(2^\Props \times \N)}^\omega$ with respect to a position $i \in \N$ and a register assignment $\regassign: \Reg \to \N$. The satisfaction relation $w,i \models_\regassign \phi$ is inductively defined as follows:
\begin{itemize}\itemsep=0.5ex
\item $w,i \models_\regassign \prop$ if $\prop \in P_i$,

\item $w,i \models_\regassign \regtest{\bowtie\,}{r}$ if $v_i \bowtie \regassign(r)$,

\item $w,i \models_\regassign \neXt \phi$ if $w,i+1 \models_\regassign \phi$,

\item $w,i \models_\regassign \Until{\phi_1}{\phi_2}$ if there is $j \ge i$ such that $w,j \models_\regassign \phi_2$ and $w,k \models_\regassign \phi_1$ for all $k \in \{i,\ldots,j-1\}$,

\item $w,i \models_\regassign \Release{\phi_1}{\phi_2}$ if one of the following holds: (i) $w,k \models_\regassign \phi_2$ for all $k \ge i$, or (ii) there is $j \ge i$ such that $w,j \models_\regassign \phi_1$ and $w,k \models_\regassign \phi_2$ for all $k \in \{i,\ldots,j\}$,

\item $w,i \models_\regassign \dfreeze{r}{\phi}$ if $w,i \models_{\regassign[r\; \mapsto\, v_i]} \phi$, where the register assignment $\regassign[r \mapsto v_i]$ maps register $r$ to $v_i$ and coincides with $\regassign$ on all other registers.
\end{itemize}
Negation, disjunction, and conjunction are interpreted as usual.

A formula $\phi$ is a \emph{sentence} when every subformula of $\phi$ of the form $\regtest{\bowtie\,}{r}$ is in the scope of a freeze quantifier $\dfreeze{r}{}$. In that case, the initial register assignment is irrelevant, and we simply write $w \models \phi$ if $w,0 \models_\regassign \phi$ (with $\regassign$ arbitrary). Let $\POCA = (Q,\params,\qinit,\Delta,\propmap)$ be an \socapc and let $\rho = (q_0,v_0) \pocatrans{\inst}  (q_1,v_1) \pocatrans{\inst} \cdots$ be an infinite run of $\POCA$. We say that $\rho$ satisfies $\phi$, written $\rho \models \phi$, if $(\propmap(q_0),v_0)(\propmap(q_1),v_1) \ldots \models \phi$. Moreover, we write $\POCA \modelsex \phi$ if there \emph{exist} $\inst$ and an infinite initialized $\inst$-run $\rho$ of $\POCA$ such that $\rho \models \phi$.

\begin{table}
\begin{center}
\belowrulesep=2mm
\begin{tabular}{rP{45mm}P{35mm}P{4cm}}
  & \textsc{Reachability}\,/\,\textsc{B\"{u}chi} & $\textsc{MC}(\LTL)$ & $\textsc{MC}(\FFLTL)$ \\[1mm] 
  \toprule
  \oca
    & \centering $\NLOGSPACE$-complete \newline~\cite{Valiant1975}\,/\!\cite{DemriG09} 
    & $\PSPACE$-complete \newline (e.g.,~\cite{Haase12})
    & \textcolor{red}{$\NEXPTIME$-complete \newline Theorem~\ref{thm:main}} \\
  \midrule
  \soca
    & \hspace{5mm}$\NP$-complete\newline\cite{HaaseKOW09}\,/\!\cite{Haase12}
    & $\PSPACE$-complete \newline~\cite{GollerHOW10}
    & \textcolor{red}{$\NEXPTIME$-complete \newline Theorem~\ref{thm:main}} \\
 \end{tabular}
\end{center}
\caption{Old and new results\label{table:results}}
\end{table}

Model checking for a class $\Class$ of \socapc and a logic $\Logic \subseteq \FLTL$ is defined as follows:
\begin{center}
\begin{decproblem}
  \problemtitle{\MC{\Class}{\Logic}}
  \probleminput{$\POCA = (Q,\params,\qinit,\Delta,\propmap)\in \Class$ and a sentence $\phi \in \Logic$}
  \problemquestion{Do we have $\POCA \modelsex \phi$\,?}
\end{decproblem}
\end{center}

Note that, following~\cite{DemriLS10,LechnerMOPW16}, we study the existential version of the model-checking problem (``Is there some run satisfying the formula?''). The reason is that the flat fragment that we consider next is not closed under negation, and for many useful \freezeLTL formulas the negation is  actually \emph{flat} (cf.\ Example~\ref{ex:fltl} below). In Section~\ref{sec:soca-universal-mc}, we will study the \emph{universal} version of the model-checking problem (``Do all runs satisfy the formula?'').

Unfortunately, \MC{\oca}{\FLTL} is undecidable~\cite{DemriLS10}, even if formulas use only one register. This motivates the study of the flat fragment of \FLTL, restricting the usage of the freeze quantifier~\cite{DemriLS10}. Essentially, it is no longer possible to overwrite a register unboundedly often.

\begin{defi}[Flat Freeze LTL]
The \emph{flat} fragment of $\FLTL$, denoted by $\FFLTL$, contains a formula $\phi$ if, for every occurrence of a subformula $\psi =\Until{\phi_1}{\phi_2}$ (respectively, $\psi = \Release{\phi_2}{\phi_1}$) in $\phi$,
 the following hold: (i) If the occurrence of $\psi$ is in the scope of an even number of negations, then the freeze quantifier does not occur in $\phi_1$, and (ii) if it is in the scope of an odd number of negations, then the freeze quantifier does not occur in $\phi_2$.
\end{defi}

\begin{exa}\label{ex:fltl}
An example \FFLTL~formula is $\Future\, \dfreeze{r}{\Globally\, \bigl((\regtest{<}{r}) \vee (\regtest{=}{r})\bigr)}$, saying that a run of an $\oca$ takes only finitely many different counter values. On the other hand, the formula $\phi = \Globally\, \dfreeze{r}{\bigl(\mathit{req} \to \Future (\mathit{serve} \wedge (\regtest{=}{r}))\bigr)}$ (from~\cite{LechnerMOPW16}), which says that every request is eventually served with the same ticket, is \emph{not} in \FFLTL, but its negation
$\neg \neg \bigl(\Until{\true}{\neg\, \dfreeze{r}{\bigl(\mathit{req} \to \Future (\mathit{serve} \wedge (\regtest{=}{r}))\bigr)}}\bigr)$ is. Indeed, the until modality lies in the scope of an even number of negations, and the
freeze quantifier is in the second argument of the until.
\end{exa}

In~\cite{DS-fossacs10}, it has been shown that \MC{\oca}{\FFLTL} can be reduced to \Nonemptiness{\textup{\ocap}} (without comparisons with constants), but decidability of the latter was left open (positive results were only obtained for a restricted fragment of $\FFLTL$). In~\cite{LechnerMOPW16}, Lechner \etal\ showed that \Nonemptiness{\textup{\ocap}} is decidable. In fact, they establish that \MC{\soca}{\FFLTL} is in $\tNEXPTIME$.

Our main result states that the problems \MC{\oca}{\FFLTL} and \MC{\soca}{\FFLTL} are both $\NEXPTIME$-complete.
A comparison with known results can be found in Table~\ref{table:results}. The proof outline is depicted in Figure~\ref{fig:outline}. Essentially, we also rely on a reduction of \MC{\soca}{\FFLTL} to \Reach{\textup{\ocap}}, and the main challenge is to establish the precise complexity of the latter. In Section~\ref{sec:poca-np}, we reduce \Reach{\textup{\ocap}} to non-emptiness of alternating two-way automata over infinite words, which is then exploited to prove an $\NP$ upper bound. The reductions from \MC{\soca}{\FFLTL} to \Reach{\textup{\ocap}} themselves are given in Section~\ref{sec:soca-mc}, which also contains our main results.

\begin{figure}
\begin{center}
\scalebox{0.9}{
\begin{gpicture}
\gasset{Nframe=y,Nh=6,Nmr=2,AHangle=25,AHLength=2,AHlength=1.9}

\node[Nw=48,Nh=41,fillcolor=gray!20] (rect) (0,-60) {}

\node[Nframe=n] (R) (0,8) {{\bf Reductions}}
\node[Nframe=n] (R) (50,8) {{\bf Complexity}}

\node[Nw=45,fillcolor=blue!20] (SOCA) (0,0) {\MC{\soca}{\FFLTL}}
\node[Nw=45,fillcolor=blue!20] (OCA) (0,-15) {\MC{\oca}{\FFLTL}}
\node[Nw=45,fillcolor=blue!20] (BPOCA) (0,-30) {\Nonemptiness{\ocap}}
\node[Nw=45,fillcolor=blue!20] (POCA) (0,-45) {\Reach{\ocap}}
\node[Nw=45,fillcolor=blue!20] (A2A) (0,-60) {\textup{A2A-}\textsc{Noemptiness}}
\node[Nw=45,fillcolor=blue!20] (BA) (0,-75) {\Nonemptiness{\textup{BA}}}

\node[Nw=45,Nh=8,fillcolor=yellow!20] (A) (50,0) {
\begin{tabular}[t] {c}
$\NEXPTIME$\\[-1ex]
\scalebox{0.8}{Corollary~4.6}
\end{tabular}}
\node[Nw=45,Nh=8,fillcolor=yellow!20] (B) (50,-15) {
\begin{tabular}[t] {c}
$\NEXPTIME$\\[-1ex]
\scalebox{0.8}{Corollary~4.4}
\end{tabular}}
\node[Nw=45,Nh=8,fillcolor=yellow!20] (C) (50,-30) {
\begin{tabular}[t] {c}
$\NP$\\[-1ex]
\scalebox{0.8}{Corollary~4.2}
\end{tabular}}
\node[Nw=45,Nh=8,fillcolor=yellow!20] (D) (50,-45) {
\begin{tabular}[t] {c}
$\NP$\\[-1ex]
\scalebox{0.8}{Theorem~3.8}
\end{tabular}}

\node[Nw=20,Nh=12,fillcolor=gray!20] (E) (38,-60) {
\begin{tabular}[t] {c}
bounded\\[-1ex]
runs\\[-0.5ex]
\scalebox{0.8}{Corollary~3.6}
\end{tabular}}
\node[Nw=20,Nh=10,fillcolor=gray!20] (F) (62,-60) {Lemma~3.7}

\drawedge[curvedepth=0,ELside=l,ELdist=1,ELdist=-7] (SOCA,OCA) {pol.~~Lemma~4.5}
\drawedge[curvedepth=0,ELside=l,ELdist=1,ELdist=-7] (OCA,BPOCA) {exp.~~[DS10]}
\drawedge[curvedepth=0,ELside=l,ELdist=1,ELpos=45,ELdist=-7] (BPOCA,POCA) {pol.~~Lemma~4.1}
\drawedge[curvedepth=0,ELside=l,ELdist=1,ELdist=-7] (POCA,A2A) {pol.~~Lemma~3.3}
\drawedge[curvedepth=0,ELside=l,ELdist=1,ELdist=-7] (A2A,BA) {exp.~~[DK08,Ser06]}

\drawedge[curvedepth=0,ELside=r,ELdist=1,linecolor=red] (rect,E) {}
\drawedge[curvedepth=0,ELside=r,ELdist=1,linecolor=red] (F,D) {}
\drawedge[curvedepth=0,ELside=r,ELdist=1,linecolor=red] (E,D) {}
\drawedge[curvedepth=0,ELside=r,ELdist=1,linecolor=red] (D,C) {}
\drawedge[curvedepth=0,ELside=r,ELdist=1,linecolor=red] (C,B) {}
\drawedge[curvedepth=0,ELside=r,ELdist=1,linecolor=red] (B,A) {}

\drawedge[curvedepth=0,AHnb=0,ELside=r,ELdist=1] (SOCA,A) {}
\drawedge[curvedepth=0,AHnb=0,ELside=r,ELdist=1] (OCA,B) {}
\drawedge[curvedepth=0,AHnb=0,ELside=r,ELdist=1] (BPOCA,C) {}
\drawedge[curvedepth=0,AHnb=0,ELside=r,ELdist=1] (POCA,D) {}

\drawline[AHnb=0,dash={1.5}0] (-39,-39.2) (85,-39.2)
\node[Nframe=n] (Section) (-32,-42) {\scalebox{0.8}{Section~3}}
\node[Nframe=n] (Section) (-32,-36) {\scalebox{0.8}{Section~4}}

\end{gpicture}
}
\end{center}
\caption{Proof structure for upper bounds\label{fig:outline}}
\end{figure}

\section{\ocap-Reachability is NP-complete}\label{sec:poca-np}

In this section, we reduce \Reach{\ocap} to the non-emptiness problem of alternating two-way automata (A2As) over infinite words.
While this already implies that \Reach{\ocap} is in $\PSPACE$,
we then go a step further and show that the problem is in $\NP$.

\subsection{From \ocaps to Alternating Two-Way Automata}

Let $\POCA = (Q,\params,\qinit,\Delta)$  be an \ocap.
The main idea is to encode a parameter instantiation $\inst:\params\to \N$ as a word over the alphabet $\Sigma = \params \cup \{\adel\}$, where $\adel$ is a fresh symbol.
A word $w = a_0 a_1 a_2 \cdots \in \Sigma^\omega$, with $a_i \in \Sigma$, is called a \emph{parameter word} (over $\params$) if $a_0 = \adel$ and, for all $\param \in \params$, there is exactly one position $i \in \N$ such that $a_i = \param$. In other words, $w$ starts with $\adel$ and every parameter occurs exactly once. Then, $w$ determines a parameter instantiation $\inst_w: \params \to \N$ as follows: if $\param = a_i$, then $\inst_w(\param) = |a_1 \cdots a_{i-1}|_\adel$ where $|a_1 \cdots a_{i-1}|_\adel$ denotes the number of occurrences of $\adel$ in $a_1 \cdots a_{i-1}$ (note that we start at the second position of $w$).
For example, given $\params=\{\param_1,\param_2,\param_3\}$, both $w = \adel x_2 \adel\adel x_1 x_3 \adel^\omega$ and $w' = \adel x_2 \adel\adel x_3 x_1 \adel^\omega$ are parameter words with
$\inst_w = \inst_{w'} = \{\param_1\mapsto 2, \param_2\mapsto 0, \param_3\mapsto 2\}$.
Note that, for every parameter instantiation $\inst$, there is at least one parameter word $w$ such that $\inst_w = \inst$.
Let $\paramwords{\params}$ denote the set of all parameter words over $\params$.

From $\POCA$ and a state $q_f \in Q$, we will build an A2A that accepts the set of parameter words $w$ such that $q_f$ is $\inst_w$-reachable. Like a Turing machine, an A2A can read a letter, change its state, and move its head to the left or to the right (or stay at the current position). In addition, it can spawn several independent copies, for example one that goes to the left and one that goes to the right. However, unlike a Turing machine, an A2A is not allowed to modify the input word so that its expressive power does not go beyond finite (B{\"u}chi) automata.  The simulation proceeds as follows. When the \ocap increments its counter, the A2A moves to the right to the next occurrence of $\adel$. To simulate a decrement, it moves to the left until it encounters the previous $\adel$. To mimic the zero test, it verifies that it is currently on the first position of the word. Moreover, it will make use of the letters in $\params$ to simulate parameter tests. At the beginning of an execution, the A2A spawns independent copies that check whether the input word is a valid parameter word.

Let us define A2As and formalize the simulation of an \ocap. Given a finite set $Y$, we denote by $\mathbb{B}^+(Y)$ the set of positive Boolean formulas over $Y$, including $\btrue$ and $\bfalse$. A subset $Y'\subseteq Y$ \emph{satisfies} $\beta \in \mathbb{B}^+(Y)$, written $Y' \models \beta$, if $\beta$ is evaluated to true when assigning $\btrue$ to every variable in $Y'$, and $\bfalse$ to every variable in $Y\backslash Y'$. In particular, we have $\emptyset \models \btrue$. For $\beta \in \mathbb{B}^+(Y)$, let $|\beta|$ denote the size of $\beta$, defined inductively by $|\beta_1 \vee \beta_2| = |\beta_1 \wedge \beta_2| = |\beta_1| + |\beta_2| + 1$ and $|\beta| = 1$ for atomic formulas $\beta$.

\begin{defi}
An \emph{alternating two-way automaton} (A2A) is a tuple $\twoafa=(S,\Sigma,s_{\mathsf{in}},\delta,S_f)$,
where
\begin{itemize}
\item $S$ is a finite set of states,
\item $\Sigma$ is a finite alphabet,
\item $s_{\mathsf{in}}\in S$ is the initial state,
\item $S_f\subseteq S$ is the set of accepting states, and
\item $\delta \subseteq S \times (\Sigma \cup \{\first\}) \times \mathbb{B}^+(S\times \{+1,0,-1\})$ is the finite transition relation\footnote{One often considers a transition \emph{function} $\delta: S \times \Sigma \times \{\first,\neg\first\} \to \mathbb{B}^+(S\times \{+1,0,-1\})$, but a relation is more convenient for us.}.
A transition $(s,\mathit{test},\beta) \in \delta$ will also be written $\afatrans{s}{\mathit{test}}{\beta}$.
\end{itemize}
The size of $\twoafa$ is defined as $|\twoafa| = |S| + |\Sigma| + \sum_{(s,\mathit{test},\beta) \in \delta} |\beta|$.
\end{defi}

While, in an \ocap, $+1$ and $-1$ are interpreted as \emph{increment} and \emph{decrement} the counter, respectively, their interpretation in an A2A is \emph{go to the right} and \emph{go to the left} in the input word. Moreover, $0$ means \emph{stay}. Actually,
when $\delta \subseteq S \times \Sigma \times (S\times \{+1\})$, then we deal with a classical \emph{one-way finite automaton}.

A \emph{run} of $\twoafa$ on an infinite word $w=a_0 a_1 a_2 \dots \in \Sigma^\omega$ is a rooted tree (possibly infinite, but finitely branching) whose vertices are labeled with elements in $S\times\N$. A node with label $(s,n)$ represents a proof obligation that has to be fulfilled starting from state $s$ and position $n$ in the input word. The root of a run is labeled by $(s_{\mathsf{in}},0)$.
Moreover, we require that, for every vertex labeled by $(s,n)$ with $k \in \N$ children labeled by $(s_1,n_1), \dots,(s_k,n_k)$, there is a transition $(s,\mathit{test},\beta) \in \delta$ such that
(i) the set $\{(s_1,n_1-n), \dots, (s_k,n_k-n)\}\subseteq S\times \{+1,0,-1\}$ satisfies $\beta$,
(ii) $\mathit{test}=\first$ implies $n=0$, and
(iii) $\mathit{test}\in\Sigma$ implies $a_n = \mathit{test}$.
Note that, similarly to an \ocap, a transition with move $-1$ is blocked if $n=0$, \ie, if $\twoafa$ is at the first position of the input word.  A run is \emph{accepting} if every infinite branch visits some accepting state from $S_f$  infinitely often.
The language of $\twoafa$ is defined as $L(\twoafa)=\{w \in \Sigma^\omega \mid \mbox{there exists an accepting run of } \twoafa \mbox{ on } w\}$. The \emph{non-emptiness problem} for A2As is to decide, given an A2A $\twoafa$, whether $L(\twoafa)\neq\emptyset$.

\begin{thmC}[\cite{Serre06}]\label{thm:2wfa}
	The non-emptiness problem for A2As	is in $\pspace$.
\end{thmC}

It is worth noting that~\cite{Serre06} also uses two-wayness to simulate one-counter automata, but in a game-based setting (the latter is reflected by alternation).

We will now show how to build an A2A from the \ocap $\POCA = (Q,\params,\qinit,\Delta)$ and a target state $q_f \in Q$.
\begin{lem}%
	\label{lemma_translation_poca_twoway}
	There is an A2A $\twoafa=(S,\Sigma,s_{\mathsf{in}},\delta,S_f)$, with $\Sigma = \params \cup \{\adel\}$, such that $L(\twoafa) = \{w \in \paramwords{\params} \mid q_f \textup{ is } \inst_w\textup{-reachable in }\POCA\}$.
    Moreover, $|\twoafa|= \mathcal{O}(|\POCA|^2)$.
\end{lem}

\begin{proof}
The states of $\twoafa$ include $Q$ (to simulate $\POCA$), a new initial state $s_{\mathsf{in}}$, and some extra states, which will be introduced below.

Starting in $s_{\mathsf{in}}$ and at the first letter of the input word, the A2A $\twoafa$ spawns several copies: $\smash{\afatrans{s_{\mathsf{in}}}{\adel}{(\qinit,0)} \wedge \bigwedge_{\param \in \params} (\search{\param},+1)}$.
The copy starting in $\qinit$ will henceforth simulate $\POCA$.
Moreover, from the new state $\search{\param}$, we will check that $\param$ occurs exactly once in the input word. This is accomplished using the transitions
$\afatrans{\search{\param}}{\param}{(\seen{\param},+1)}$, as well as
$\afatrans{\search{\param}}{y}{(\search{\param},+1)}$ and
$\afatrans{\seen{\param}}{y}{(\seen{\param},+1)}$ for all $y \in \Sigma \setminus \{\param\}$. Thus, $\seen{\param}$ signifies that $\param$ has been seen and must not be encountered again.
Since the whole word has to be scanned, $\seen{\param}$ is visited infinitely often so that we set $\seen{\param} \in S_f$.

It remains to specify how $\twoafa$ simulates $\POCA$. The underlying idea is very simple. A configuration $(q,v) \in \Conf{\POCA}$ of $\POCA$ corresponds to the configuration/proof obligation $(q,i)$ of $\twoafa$ where position $i$ is the $(v+1)$-th occurrence of $\adel$ in the input parameter word. The simulation then proceeds as follows.
\begin{description}\itemsep=1ex
\item[Increment/decrement] To mimic a transition $(q,+1,q') \in \Delta$,
the A2A $\twoafa$ simply goes to the next occurrence of $\adel$ on the right hand side and enters $q'$. This is accomplished by several A2A-transitions: $\smash{\afatrans{q}{\adel}{(\gotoright{q'},+1)}}$, $\afatrans{\gotoright{q'}}{\param}{(\gotoright{q'},+1)}$ for every $\param \in \params$, and $\smash{\afatrans{\gotoright{q'}}{\adel}{(q',0)}}$.  Decrements are handled similarly, using states of the form $\gotoleft{q'}$. Finally, $(q,0,q') \in \Delta$ is translated to $\smash{\afatrans{q}{\adel}{(q',0)}}$.

\item[Equality test] A zero test $(q,\zero,q') \in \Delta$ corresponds to $\afatrans{q}{\first}{(q',0)}$. To simulate a transition $(q,\ptest{=}{\param},q') \in \Delta$, the A2A spawns two copies. One goes from $q$ to $q'$ and stays at the current position. The other goes to the right and accepts if it sees $\param$ before hitting another $\adel$-symbol. Thus, we introduce a new state $\present{\param}$ and transitions $\smash{\afatrans{q}{\adel}{(q',0) \wedge (\present{\param},+1)}}$, $\afatrans{\present{\param}}{x}{\btrue}$, and $\afatrans{\present{\param}}{y}{(\present{\param},+1)}$ for all $y \in \params \setminus \{x\}$.
Note that there is \emph{no} transition that allows $\twoafa$ to read $\adel$ in state $\present{\param}$.

\item[Inequality tests] To simulate a test of the form $\ptest{>}{\param}$, we proceed similarly to the previous case. The A2A generates a branch in charge of verifying that the parameter $\param$ lies strictly to the left of the current position. Note that transitions with label $\ptest{<}{\param}$ are slightly more subtle, as $\param$ has to be retrieved strictly \emph{beyond} the next delimiter $\adel$ to the right of the current position. More precisely, $(q,\ptest{<}{\param},q') \in \Delta$ translates to $\smash{\afatrans{q}{\adel}{(q',0) \wedge (\searchn{\param},+1)}}$. We stay in $\searchn{\param}$ until we encounter the next occurrence of $\adel$. We then go into $\search{\param}$ (introduced at the beginning of the proof), which will be looking for an occurrence of $\param$. Formally, we introduce $\afatrans{\searchn{\param}}{y}{(\searchn{\param},+1)}$ for all $y \in \params \setminus \{\param\}$, and $\smash{\afatrans{\searchn{\param}}{\adel}{(\search{\param},+1)}}$.
\end{description}

\noindent
In state $q_f$, we accept: $\afatrans{q_f}{\adel}{\btrue}$. The only infinite branches in an accepting run are those that eventually stay in a state of the form $\seen{\param}$.
Thus, we set $S_f = \{\seen{\param}\mid\param\in\params\}$. It is not hard to see that $L(\twoafa) = \{w \in \paramwords{\params} \mid q_f \textup{ is } \inst_w\textup{-reachable}\}$.
Note that $\twoafa$ has a linear number of states and a quadratic number of transitions (some transitions of $\POCA$ are simulated by $\mathcal{O}(|\params|)$-many transitions of $\twoafa$).
\end{proof}

A similar reduction takes care of B{\"u}chi reachability.
Together with Theorem~\ref{thm:2wfa}, we thus obtain the following:

\begin{cor}%
\label{corrollary-pspace}
\Reach{\textup{\ocap}} and \Nonemptiness{\textup{\ocap}} are both in $\pspace$.
\end{cor}

Note that we are using alternation only to a limited extent. We could also reduce reachability in \ocaps to non-emptiness of the intersection of several two-way automata. However, this would require a more complicated word encoding of parameter instantiations, which then have to include letters of the form $\ptest{<}{\param}$ and $\ptest{>}{\param}$.

\subsection{\ocap-Reachability is in NP}

We will now show that we can improve the upper bounds given in Corollary~\ref{corrollary-pspace} to $\NP$.
This upper bound is then optimal: $\NP$-hardness, even in the presence of a single parameter, can be proved using a straightforward reduction from the non-emptiness problem for nondeterministic two-way automata over finite words over a \emph{unary} alphabet and two end-markers, which is $\NP$-complete~\cite{Galil76}.

In a first step, we exploit our reduction from \ocaps to A2As to establish a bound on the parameter values. To solve the reachability problems, it will then be sufficient to consider parameter instantiations up to that bound. For Lemma~\ref{lem:bound} and Corollary~\ref{cor:bound} below, we fix an \ocap $\POCA=(Q,\params,\qinit,\Delta)$ and a state $q_f\in Q$.

\begin{lem}\label{lem:bound}
There is $\bound \in 2^{\mathcal{O}(|\POCA|^4)}$ such that the following holds:
If $\Reachpred{\POCA}{q_f}$, then there is a parameter instantiation $\inst: \params \to \N$ such that $\inst(\param) \le \bound$ for all $\param \in \params$ and $q_f$ is $\inst$-reachable in $\POCA$.
\end{lem}

\begin{proof}
According to Lemma~\ref{lemma_translation_poca_twoway},
there is an A2A $\twoafa=(S,\params \cup \{\adel\},s_{\mathsf{in}},\delta,S_f)$ such that
$L(\twoafa) = \{w \in \paramwords{\params} \mid q_f \textup{ is } \inst_w\textup{-reachable in }\POCA\}$ and $|\twoafa|= \mathcal{O}(|\POCA|^2)$. By~\cite{Vardi98}, there is a (nondeterministic) B{\"u}chi automaton $\B$ such that $L(\B) = L(\twoafa)$ and $|\B| \in 2^{\mathcal{O}(|\twoafa|^2)}$ (cf.\ also~\cite{DaxK08}).

Let $\bound = |\B|$. Suppose $\Reachpred{\POCA}{q_f}$. This implies that $L(\B) \neq \emptyset$. But then, there must be a word $u \in \Sigma^\ast$ such that $|u| \le |\B|$ ($|u|$ denoting the length of $u$) and $w=u\adel^\omega \in L(\B)$. We have that $q_f$ is $\inst_{w}$-reachable and $\inst_{w}(\param) \le |\B| = \bound$ for all $\param \in \params$.
\end{proof}

As a corollary, we obtain that it is sufficient to consider bounded runs only.
For a parameter instantiation $\inst: \params \to \N$ and a bound $\bound \in \N$, we say that a $\inst$-\emph{run} $\rho = (q_0,v_0) \pocatrans{\inst}  (q_1,v_1) \pocatrans{\inst}  (q_2,v_2) \pocatrans{\inst} \cdots$ is $\bound$-\emph{bounded} if all its counter values $v_0,v_1,\ldots$ are at most $\bound$.

\begin{cor}\label{cor:bound}
There is $\bound \in 2^{\mathcal{O}(|\POCA|^4)}$ such that the following holds:
If $\Reachpred{\POCA}{q_f}$, then there is a parameter instantiation $\inst: \params \to \N$ such that
$\inst(\param) \le \bound$ for all $\param \in \params$ and $q_f$ is reachable within a $\bound$-bounded $\inst$-run.
\end{cor}

\begin{proof}
Consider $\bound \in 2^{\mathcal{O}(|\POCA|^4)}$ due to Lemma~\ref{lem:bound}. A standard argument in \ocas with $n$ states is that, for reachability, it is actually sufficient to consider runs up to some counter value in $\mathcal{O}(n^3)$~\cite{LLT05,chistikov-shortest-16}. We can apply the same argument here to deduce, together with Lemma~\ref{lem:bound}, that parameter and counter values can be bounded by $\bound + \mathcal{O}(|Q|^3)$.
\end{proof}

The algorithm that solves \Reach{\textup{\ocap}} in $\NP$ will guess a parameter instantiation $\inst$ and a maximal counter value in binary representation (thus, of polynomial size). It then remains to determine, in polynomial time, whether the target state is reachable under this guess. To this end, we will exploit the lemma below due to Galil~\cite{Galil76}.

Let $\POCA = (Q,\qinit,\Delta)$ be an \oca, $q,q' \in Q$, and $v,v' \in \N$. A run $(q_0,v_0) \pocatrans{}(q_1,v_1) \pocatrans{}  \cdots  \pocatrans{}(q_{n-1},v_{n-1}) \pocatrans{} (q_n,v_n)$ of $\POCA$, $n \in \N$, is called a $(v,v')$-\emph{run} if $\min\{v,v'\} < v_i < \max\{v,v'\}$ for all $i \in \{1,\ldots,n-1\}$. In other words,
all intermediate configurations have counter values strictly between $v$ and $v'$.

\begin{lemC}[\cite{Galil76}]\label{thm:galil}%
	\label{lemma_galil}
The following problems can be solved in polynomial time:\\[0.5ex]
\textup{
\begin{tabular}{ll}
\normalsize\textup{\bf Input:} & \normalsize An \oca $\POCA = (Q,\qinit,\Delta)$, $q,q' \in Q$, and $v,v' \in \N$ given in binary.\\[0.2ex]
\normalsize \textup{\bf Question 1:} & \normalsize Is there a $(v,v')$-run from $(q,v)$ to $(q',v')$\,?\\[0.2ex]
\normalsize \textup{\bf Question 2:} & \normalsize Is there a $(v,v')$-run from $(q,v)$ to $(q',v)$\,?
\end{tabular}}
\end{lemC}

\noindent
Actually,~\cite{Galil76} uses the polynomial-time algorithms stated in Lemma~\ref{lemma_galil} to show that non-emptiness of two-way automata on finite words over a unary alphabet is in $\NP$ (cf.\ the proof of the corresponding lemma on page 84 of~\cite{Galil76}, where the end-markers of a given word correspond to $v$ and $v'$).

We can now deduce the following result, which is an important step towards model checking, but also of independent interest.

\begin{thm}\label{thm:reachNP}
	\Reach{\textup{\ocap}} is $\NP$-complete.
\end{thm}

\begin{proof}[Sketch of proof] Let $\POCA=(Q,\params,\qinit,\Delta)$ be an \ocap with $\params=\{\param_1,\dots,\param_n\}$, and let $q_f\in Q$ be a state of $\POCA$. Without loss of generality, we consider reachability of the \emph{configuration} $(q_f,0)$ (since one can add a new target state and a looping decrement transition).

	Our nondeterministic polynomial-time algorithm proceeds as follows. First, it guesses $\bound \in 2^{\mathcal{O}(|\POCA|^4)}$ due to Corollary~\ref{cor:bound} (note that the binary representation of $\bound$ is of polynomial size with respect to $|\POCA|$), as well as some parameter instantiation $\inst$ satisfying $\inst(\param_i)\le \bound$ for all $1\leq i \leq n$ where each $\bound_i \df \inst(\param_i)$ is represented in binary, too. We may assume $n \ge 1$ and that $d_0<d_1<d_2<\cdots<d_n<d_{n+1}$, where $d_0=0$ and $d_{n+1}=d$ (otherwise, we can rename the parameters accordingly). Second, the algorithm checks that there exists a $\bound$-bounded $\inst$-run of the form
	\[(q_0,v_0)\pocatrans{\inst}^* (q'_0,v_0)\pocatrans{\inst}^* (q_1,v_1)\pocatrans{\inst}^* (q'_1,v_1)\pocatrans{\inst}^*\cdots\pocatrans{\inst}^*(q_k,v_k)\pocatrans{\inst}^*(q'_k,v_k)\]
	such that (letting $D = \{d_0,d_1,\ldots,d_n,d_{n+1}\}$)
	\begin{enumerate}[label=\textnormal{(\arabic*)}]
	\item $(q_0,v_0)=(\qinit,0)$ and $(q'_k,v_k)=(q_f,0)$,
	\item $v_j\in D$ for all $j \in \{0,\ldots,k\}$,
	\item\label{it:constant} between $(q_j,v_j)$ and $(q'_j,v_j)$, the counter values always equal $v_j$, for all $j \in \{0,\ldots,k\}$,
	\item\label{it:between} (strictly) between $(q'_j,v_j)$ and $(q_{j+1},v_{j+1})$, the counter values are always different from the values in $D$, for all $j \in \{0,\ldots,k-1\}$.
	\end{enumerate}
Note that we can assume $k \leq |Q|\cdot(|\params|+2)$, since in every longer run, the exact same configuration is encountered at least twice.

Hence, to check whether there exists a finite initialized $d$-bounded $\inst$-run $(\qinit,0)\pocatrans{\inst}^* (q_f,0)$, it is sufficient to guess $2k$ configurations, where  $k \leq |Q|\cdot(|\params|+2)$, and to verify that these configurations contribute to constructing a run of the form described above. This can be checked in polynomial time: The case (3) is obvious, and (4) is due to Lemma~\ref{lemma_galil}.

\medskip

The lower bound can be easily obtained from~\cite{Galil76} stating that the non-emptiness problem for nondeterministic two-way automata over finite words over a \emph{unary} alphabet and two end-markers is $\NP$-complete.%
\footnote{On the other hand, the non-emptiness problem for nondeterministic two-way automata over a unary alphabet and with a single end-marker (which can be simulated by an \oca) is $\NLOGSPACE$-complete.}
In fact, we can easily build an  \ocap with a single parameter  that simulates the behavior of such an automaton and where the parameter is used to guess the length of  the read word (and, consequently, to detect the end-marker).
\end{proof}

\subsection{Dealing with binary constants}

The previous $\NP$ algorithm can be adapted to the case of \ocapc where, in addition to parameters, the counter values can be compared with constants $c\in\N$ encoded in binary. To this end, we transform a given \ocapc $\POCA$ into an $\ocap$ $\POCA'$ by interpreting every constant $c$ from $\POCA$ as a parameter $x_c$ and build the  corresponding A2A as stated by Lemma~\ref{lemma_translation_poca_twoway}. Then, we can adapt the proof of Lemma~\ref{lem:bound} to show that, if $\Reachpred{\POCA}{q_f}$ for a given $q_f$, then there is a parameter instantiation where all parameter values can be bounded by an exponential bound. This is  enough to obtain the desired  upper-bound following the same reasoning as in the proof of Theorem~\ref{thm:reachNP}.

\begin{thm}\label{thm:creachNP}
	\Reach{\textup{\ocapc}} is $\NP$-complete.
\end{thm}

\begin{proof}
Let $\POCA=(Q,\params,\qinit,\Delta)$ be an \ocapc and let $q_f\in Q$. Moreover, let $C=\{c \in \mathbb{N} \mid c \mbox{ appears in } \cTrans\}$. From $\POCA$, we build an \ocap  $\POCA'=(Q,\params',\qinit,\Delta')$ where $\params'=\params\cup \{x_c \mid c \in C\}$ and $\Delta'$ is  obtained from $\Delta$ by replacing each transition $(q,\bowtietest{c},q') \in  \cTrans$ by $(q,\bowtietest{x_c},q')$  and keeping the other transitions.

Now, using Lemma~\ref{lemma_translation_poca_twoway}, we  know there is an A2A $\twoafa'$ such that $L(\twoafa') = \{w \in \paramwords{\params'} \mid q_f \textup{ is } \inst_w\textup{-reachable in } \POCA'\}$ and $|\twoafa'|= \mathcal{O}(|\POCA'|^2)$. As in the proof of Lemma~\ref{lem:bound}, we know then from~\cite{Vardi98} that there is a (nondeterministic) B{\"u}chi automaton $\B$ such that $L(\B) = L(\twoafa')$ and $|\B| \in 2^{\mathcal{O}(|\twoafa'|^2)}$. Suppose $\Reachpred{\POCA}{q_f}$. This implies that there is a parameter word $w$ over $\params'$ that belongs to $L(\B)$ and such that $\inst_w(x_c)=c$ for all $c \in C$.
Let $c_{\max}$ be maximal in $C$. Note that, since the constants are encoded in binary in $\POCA$, we have $c_{\max}\in 2^{\mathcal{O}(|\POCA|)}$.
Then, there must also be a word $u \in {(\params' \cup \{\adel\})}^\ast$ such that $|u| \le |\B| + c_{\max}$ ($|u|$ denoting the length of $u$), $w' \df u\adel^\omega \in L(\B)$, and $\inst_{w'}(x_c)=c$ for all $c \in C$. Hence, $q_f$ is $\inst_{w'}$-reachable and $\inst_{w'}(\param) \le c_{\max}+|\B| =: \bound$ for all $\param \in \params'$. Note that $d \in 2^{\mathcal{O}(|\POCA|^4)}$.

Then, we follow the same reasoning as in the proof of Theorem~\ref{thm:reachNP}:
We guess a parameter instantiation $\gamma$ satisfying $\gamma(x) \le d$ for all $x \in \params'$.
In fact, the only difference is that, for $c \in C$, we do not \emph{guess} $\gamma(x_c)$ but \emph{determine} $\gamma(x_c)$ to be $c$.
This ensures that, if we find a $\gamma$-run in $\POCA'$ reaching $q_f$, then $\Reachpred{\POCA}{q_f}$. Furthermore, the previous reasoning ensures that the bound $d$ is enough to witness a run in $\POCA$ reaching $q_f$.  This allows us to deduce that	\Reach{\textup{\ocapc}} is in $\NP$. The hardness part comes from the fact that the problem is already $\NP$-hard when considering \ocap.
\end{proof}

\begin{rem}
 Using a reduction from~\cite{BundalaO17}, Theorem~\ref{thm:creachNP} will allow us to establish an optimal $\NEXPTIME$ algorithm for deciding non-emptiness of closed parametric timed automata with a single parametric clock~\cite{AlurHV93,BundalaO17} (cf.\ Section~\ref{sec:pta}).
\end{rem}

\section{From \ocapc-Reachability to \soca-Model-Checking}\label{sec:soca-mc}

This section is dedicated to model checking \ocas against \FFLTL. In Section~\ref{ref:upper-blounds}, we establish a couple of reductions from \MC{\soca}{\FFLTL} down to \Reach{\textup{\ocap}}, which allow us to conclude that the former problem is in $\NEXPTIME$. Then, in Section~\ref{ref:lower-blounds}, we provide a matching lower bound. 

\subsection{Upper Bounds}\label{ref:upper-blounds}

First, we use the fact that \Reach{\textup{\ocap}} is in $\NP$ (Theorem~\ref{thm:reachNP}) to show that \Nonemptiness{\textup{\ocap}} is in $\NP$, too:

\begin{figure}[t]
\centering
\scalebox{.74}{\begin{tikzpicture}[->,>=stealth',shorten >=1pt,auto,node distance=4cm,
thick,node/.style={circle,draw,scale=0.9},
 roundnode/.style={circle, draw=black,  thick, minimum size=5mm},]
\tikzset{every state/.style={minimum size=0pt}};

\node[roundnode, initial, initial text={}] (si) at (0,0) {};
\node(silabel) at (0,0) {$\qinit$};

\node[roundnode] (aux1) at (1.5,1) {};
\node[roundnode] (aux2) at (1.5,-1) {};
\node at (1.5,-1) {$q$};


\node[roundnode] (r) at (3,0.7) {};
\node(rlabel) at (3,0.7) {$\newr$};
\node(dots1) at (3,-0.3) {\LARGE{$\dots$}};
\node[roundnode] (aux3) at (4.5,0.5) {};

\node[roundnode] (s) at (6,0) {};
\node(slabel) at (6,0) {$q_f$};
\node[roundnode] (p) at (8,0) {};
\node(plabel) at (8,0) {$\newp$};

\node[roundnode] (sip) at (9.5,0) {};
\node(siplabel) at (9.5,0) {$\hat{q}_\mathsf{in}$};

\node[roundnode] (aux1p) at (11,1) {};
\node[roundnode] (aux2p) at (11,-1) {};
\node at (11,-1) {$\hat{q}$};
\node[roundnode] (rp) at (12.5,0.7) {};
\node(rplabel) at (12.5,0.7) {$\hat\newr$};
\node(dots1) at (12.5,-0.3) {\LARGE{$\dots$}};
\node[roundnode] (sp) at (15.5,0) {};
\node[roundnode] (aux3p) at (14,0.5) {};
\node(splabel) at (15.5,0.01) {$\hat{q}_f$};
\node[roundnode] (q) at (17.5,0) {};
\node(qlabel) at (17.5,0) {$\news$};

\node[roundnode] (r1) at (9.5,2) {};
\node(r1label) at (9.5,2) {$\newr_1$};

\node[roundnode] (r2) at (11.5,2) {};
\node(r2label) at (11.5,2) {$\newr_2$};

\node(dots) at (13.5,2) {$\dots$};

\node[roundnode] (rn) at (15.5,2) {};
\node(rn-1label) at (15.5,2) {$\newr_{n}$};

\path[->] (r) edge [bend left=10]  node [near end,sloped] {\scriptsize{$0$}} node [below,near end,sloped,xshift=-3mm] {} (r1);

\path[->] (r1) edge  node [midway] {\scriptsize{$\ptest{>}{\param_1}$}} node [below,midway] {}  (r2);
\path[->] (rn) edge [bend left=25]  node [near start,sloped] {\scriptsize{$\ptest{>}{\param_n}$}}  (q);

\draw (-0.8,1.5) rectangle (6.5,-1.5);

\draw (9,1.5) rectangle (16,-1.5);

\path[->, draw=orange,very thick] (s) edge  node [midway] {\scriptsize{$\ptest{=}{\newxp}$}} node [below,midway] {} (p);

\path[->, draw=orange,very thick] (sp) edge  node [midway] {\scriptsize{$\ptest{=}{\newxp}$}} node [below,midway] {}  (q);

\path[->, draw=teal,very thick] (si) edge  node [midway,sloped] {} (aux1);
\path[->, draw=teal,very thick] (s) edge [bend left=25]  node [midway,sloped] {} (aux2);
\path[->, draw=teal,very thick] (sp) edge [bend left=25]  node [midway,sloped] {} (aux2p);

\path[->, draw=magenta,very thick] (p) edge [bend right=25]  node [midway,sloped,below] {} (aux2p);

\path[->, draw=teal,very thick] (sip) edge  node [midway,sloped,below] {} (aux1p);


\path[->, draw=teal,very thick] (aux3) edge  node [midway,sloped] {} (s);
\path[->, draw=teal,very thick] (aux3p) edge  node [midway,sloped,below] {} (sp);


	\end{tikzpicture}

}
\caption{The \ocap $\POCA'$ constructed from $\POCA$ in the proof of Lemma~\ref{lem:buechi2reach}.
}%
\label{fig:poca_repeated_reach}
\end{figure}

\begin{lem}%
	\label{lem:buechi2reach}
	\Nonemptiness{\textup{\ocap}} is polynomial-time reducible to \Reach{\textup{\ocap}}.
\end{lem}

\begin{proof}
	Let $\POCA=(Q,\params,\qinit,\Delta)$ be an \ocap with $\params=\{\param_1,\dots,\param_n\}$ (without loss of generality, we suppose $n \ge 1$), and let $F\subseteq Q$.
	It is sufficient to give an algorithm that determines whether a particular state $q_f\in F$ can be repeated infinitely often. Thus, we construct an \ocap $\POCA'$, with a distinguished state $\news$, such that
	$\RepReachpred{\POCA}{q_f}$ iff $\Reachpred{\POCA'}{\news}$.

	To this end, we first define an \oca $\OCA$, which is obtained from $\POCA$ by (i) removing all transitions whose labels are of the form $\zero$, $\ptest{=}{\param}$, or $\ptest{<}{\param}$, for some $\param\in\params$, and (ii) replacing all transition labels $\ptest{>}{\param}$ (for some $\param\in\params$) by $0$.
	Then, we compute the set $\newR\subseteq Q$ of states $\newr$ of $\OCA$ from which there is an infinite run starting in $(\newr,0)$ and visiting $q_f$ infinitely often. This can be done in $\NLOGSPACE$~\cite{DemriG09}.

	Now, we obtain the desired \ocap $\POCA'=(Q',\params',\qinit,\Delta')$ as follows. The set of states is $Q'= Q \uplus \{\hat{q} \mid q\in Q\} \uplus \{\newp,\news,\newr_1,\dots,\newr_{n}\}$. That is, we introduce a copy $\hat{q}$ for every $q \in Q$ as well as fresh states $\newp,\news,\newr_1, \dots \newr_{n}$.
Recall that $\news$ will be the target state of the reachability problem we reduce to. Moreover, $\params' = \params \uplus \{\newxp\}$ where $\newxp$ is a fresh  parameter.
It remains to define the transition relation $\Delta'$, which includes $\Delta$ as well as some new transitions (\cf~Figure~\ref{fig:poca_repeated_reach}). Essentially, there are two cases to consider:

(1) First, $q_f$ may be visited infinitely often in $\POCA$, and infinitely often along with the same counter value. To take this case into account, we use the new parameter $\newxp$ and a new transition $(q_f,\ptest{=}{\newxp},\newp)$, which allows $\POCA'$ to ``store'' the current counter value in $\newxp$. From $s$, we then enter and simulate the copy of $\POCA$, \ie, we introduce transitions $(\newp,\op,\hat{q})$ for all $(q_f,\op,q) \in \Delta$, as well as $(\hat{q}_1,\op,\hat{q}_2)$ for all $(q_1,\op,q_2)\in \Delta$. If, in the copy, $\hat{q}_f$ is visited along with the value stored in $\newxp$, we may thus enter $\news$.

(2) Now, suppose that a $\inst$-run $\rho$ of $\POCA$ visits $q_f$ infinitely often, but not infinitely often with the same counter value. If $\rho$ visits some counter value $v \le \max\{\inst(x) \mid x \in \params\}$ infinitely often, then it necessarily contains a subrun of the form $(q,v) \pocatrans{\inst}^* (q_f,v') \pocatrans{\inst}^* (q,v)$ for some $q\in Q$ and $v' \in \N$. But then, there is an infinite $\inst$-run that visits $(q_f,v')$ infinitely often so case (1) above applies. Otherwise, $\rho$ will eventually stay strictly above $\max\{\inst(x) \mid x \in \params\}$.
Thus, we add the following transitions to $\Delta'$, which will allow $\POCA'$ to move from $\newr \in \newR$ to $\news$ provided the counter value is greater than the maximal parameter value: $(\newr,0,\newr_1)$ for all $\newr \in \newR$, as well as $(\newr_1,\ptest{>}{\param_1},\newr_2),(\newr_2,\ptest{>}{\param_2},\newr_3),\dots,(\newr_{n},\ptest{>}{\param_n},\news)$.
\end{proof}

From Theorem~\ref{thm:reachNP} and Lemma~\ref{lem:buechi2reach}, we obtain the exact complexity of
\Buechi{\textup{\ocap}}
(the lower bound is by a straightforward reduction from \Reach{\textup{\ocap}}).

\begin{cor}%
    \label{cor:buechi-np}
	\Buechi{\textup{\ocap}} is $\NP$-complete.
\end{cor}

The following link between \MC{\textup{\oca}}{\FFLTL} and \Nonemptiness{\textup{\ocap}} is due to~\cite{DS-fossacs10}:

\begin{lemC}[\cite{DS-fossacs10}]%
	\label{lem:mc2ocap}
Let $\POCA$ be an \oca and $\phi \in \FFLTL$ be a sentence.
One can compute, in exponential time, an \ocap $\POCA' = (Q,\params,\qinit,\Delta)$ (of exponential size) and a set $F \subseteq Q$ such that
$\POCA \modelsex \phi$ iff $\RepReachpred{\POCA'}{q_f}$ for some $q_f \in F$.
\end{lemC}

Actually, the construction from~\cite{DS-fossacs10} only considers the fragment of $\FFLTL$ without any register tests of the form $\regtest{<}{r}$ and $\regtest{>}{r}$, but it can easily be extended to handle them.
On the other hand, the restriction to \FFLTL~is crucial. It allows one to assume that a register is written at most once so that, in the \ocap, it can be represented as a parameter. In particular, $\params$ is the set of registers that occur in $\phi$. From Corollary~\ref{cor:buechi-np} and Lemma~\ref{lem:mc2ocap}, we now deduce our first model-checking result:

\begin{cor}\label{poca-nonempt-np}
	\MC{\textup{\oca}}{\FFLTL} is in $\NEXPTIME$.
\end{cor}


However, it turns out that \soca model checking is no harder than \oca model checking. The reason is that succinct updates can be encoded in small LTL formulas.

\begin{lem}%
	\label{thm:ltl_succinct_poca_to_poca}
	\MC{\textup{\soca}}{\FFLTL} is polynomial-time reducible to \MC{\textup{\oca}}{\FFLTL}.
\end{lem}

\begin{proof}
Let $\POCA = (Q,\qinit,\Delta,\propmap)$ be an \soca and $\phi \in \FFLTL$ be a sentence.
Without loss of generality, we assume that $\phi$ is in negation normal form, \ie,
negation is applied only to atomic propositions. This is thanks to the logical equivalences $\neg(\Until{\phi_1}{\phi_2}) \equiv \Release{\neg\phi_1}{\neg\phi_2}$ and $\neg(\Release{\phi_1}{\phi_2}) \equiv \Until{\neg\phi_1}{\neg\phi_2}$.
We construct an \oca $\POCA' = (Q',\qinit',\Delta',\propmap')$ and a sentence $\phi' \in \FFLTL$ of polynomial size such that
$\POCA \modelsex \phi$ iff $\POCA' \modelsex \phi'$.

Let $Z = \{z \mid (q,z,q') \in \Delta \cap (Q \times \Z \times Q)$ with $|z| \ge 2\}$ and let
$\Lambda = \{\bcsep{z} \mid z \in Z\} \cup \{\zerobit,\onebit\}$ be a set of fresh propositions.
To obtain $\POCA'$ from $\POCA$, we replace every transition $(q,z,q') \in \Delta \cap (Q \times Z \times Q)$ by the gadget depicted in Figure~\ref{fig:nfa_binary_counting}. The gadget implements a binary counter generating sequences (of sets of propositions) of the form depicted in Figure~\ref{fig:countingseq}, for $z=6$. We assume that the least significant bit is on the left. To make sure that the binary counter works as requested, we define an LTL formula  $\mathit{Counter}$.

\begin{figure}[t]
\centering
\scalebox{.74}{\begin{tikzpicture}[>=latex',shorten >=1pt,node distance=1.5cm,on grid,auto, thick,
roundnode/.style={circle, draw=black, thick, minimum size=6mm},
squarenode/.style={rectangle, draw,dashed,minimum size=2mm},
ellipsenode/.style={draw, rounded rectangle, draw=black!60, very thick, minimum size=5mm}
]
\node[roundnode] (q) at (0,0) {};
\node at (0,0) {$q$};

\node[roundnode] (hashtag)  [right =1.5cm of q]  {};
\node [squarenode] [above=.8cm of hashtag] {$\bcsep{z}$};

\node[roundnode] (bit0)  [right =1.5cm of hashtag, yshift=1cm]  {};
\node[roundnode] (bit0c)  [right =1.5cm of hashtag, yshift=-1cm]  {};

\node [squarenode] [above=.8cm of bit0] {$\onebit$};
\node [squarenode] [below=.8cm of bit0c] {$\zerobit$};

\node (hashtagdummytop)  [above =1cm of bit0]  {};
\node (hashtagdummybot)  [below =1cm of bit0c]  {};

\node [right =1.5cm of hashtag, yshift=1cm]  {$1$};
\node [right =1.5cm of hashtag, yshift=-1cm]  {$1'$};

\node[roundnode] (bit1)  [right =1.5cm of bit0]  {};
\node[roundnode] (bit1c)  [right =1.5cm of bit0c]  {};
\node [squarenode] [above=.8cm of bit1] {$\onebit$};
\node [squarenode] [below=.8cm of bit1c] {$\zerobit$};

\node[right =1.5cm of bit0]  {$2$};
\node [right =1.5cm of bit0c]  {$2'$};

\node[roundnode] (bit2)  [right =1.5cm of bit1]  {};
\node[roundnode] (bit2c)  [right =1.5cm of bit1c]  {};
\node[right =1.5cm of bit1]  {$3$};
\node [right =1.5cm of bit1c]  {$3'$};

\node [squarenode] [above=.8cm of bit2] {$\onebit$};
\node [squarenode] [below=.8cm of bit2c] {$\zerobit$};

\node (bit3)  [right =1.5cm of bit2]  {$\dots$};
\node (bit3c)  [right =1.5cm of bit2c]  {$\dots$};

\node[roundnode] (bitN)  [right =1.5cm of bit3]  {};
\node[roundnode] (bitNc)  [right =1.5cm of bit3c]  {};
\node[right =1.5cm of bit3]  {$n$};
\node[right =1.5cm of bit3c]  {$n'$};

\node[roundnode]  (hashtag2) [right =1.5cm of bitN,yshift=-1cm] {};

\node [squarenode] [above=.8cm of hashtag2] {$\bcsep{z}$};

\node[roundnode]  (q') [right =1.5cm of hashtag2] {};
\node [right =1.5cm of hashtag2] {$q'$};

\node [squarenode] [above=.8cm of bitN] {$\onebit$};
\node [squarenode] [below=.8cm of bitNc] {$\zerobit$};

\path [->] (q) edge (hashtag);
\path [->] (hashtag) edge (bit0);
\path [->] (hashtag) edge (bit0c);

\path [->] (bit0) edge (bit1);
\path [->] (bit0) edge (bit1c);
\path [->] (bit0c) edge (bit1);
\path [->] (bit0c) edge (bit1c);

\path [->] (bit1) edge (bit2);
\path [->] (bit1) edge (bit2c);
\path [->] (bit1c) edge (bit2);
\path [->] (bit1c) edge (bit2c);

\path [->] (bit2) edge (bit3);
\path [->] (bit2) edge (bit3c);
\path [->] (bit2c) edge (bit3);
\path [->] (bit2c) edge (bit3c);

\path [->] (bit3) edge (bitN);
\path [->] (bit3) edge (bitNc);
\path [->] (bit3c) edge (bitN);
\path [->] (bit3c) edge (bitNc);

\path [->] (hashtag2) edge (q');

\path [->] (bitN) edge[cyan] node [midway,above] {\scriptsize{$\pm 1$}} (hashtag2);

\path [->] (hashtag2) edge[out=170,in=330] (bit0);
\path [->] (hashtag2) edge[out=190,in=30] (bit0c);

\path [->] (bitNc) edge[cyan] node [midway,below] {\scriptsize{$\pm 1$}}(hashtag2);

\end{tikzpicture}

}
\caption{The \oca-gadget for simulating an \soca-transition $(q,z,q')$ with binary update $z$. The new propositions are depicted in dashed boxes.
States $1, \dots, n,1', \dots, n'$, where $n=\mathit{bits}(z)$, represent the bits needed to encode $z$ in binary. At the transitions originating from $n$ and $n'$, the counter is updated by $+1$ or $-1$, depending on whether $z$ is positive or negative, respectively.}%
\label{fig:nfa_binary_counting}
\end{figure}

\begin{figure}[t]
\begin{center}
\begin{align*}
&{\propmap(q)} \,\bcsep{6}\,  \onebit \zerobit \zerobit \,\,\bcsep{6}\, \zerobit \onebit \zerobit \,\bcsep{6}\, \onebit \onebit \zerobit \,\bcsep{6}\, \zerobit \zerobit \onebit \,\bcsep{6}\, \onebit \zerobit \onebit \,\bcsep{6}\, \zerobit \onebit \onebit \,\bcsep{6}\, \propmap(q')\\[-0.6ex]
&\hspace{0.6em} \uparrow \hspace{6.6em} \uparrow \hspace{11.2em} \uparrow \hspace{2.9em} \uparrow\\[-0.6ex]
&\hspace{-0.8em}\firstsep(\bcsep{6}) \hspace{3.7em} \eqsuff{\bcsep{6}} \hspace{7.3em}\lastsep^-(\bcsep{6})  \hspace{1em}\lastsep(\bcsep{6})
\end{align*}
\end{center}
\caption{A counting sequence\label{fig:countingseq}}
\end{figure}

Towards its definition, let us first introduce some notation and abbreviations. For $z \in Z$, let $\mathit{bits}(z)$ denote the number of bits needed to represent the binary encoding of $|z|$. For example, $\mathit{bits}(6) = 3$. For $i \in \{1,\ldots,\mathit{bits}(z)\}$, we let $\mathit{bit}_i(z)$ denote the $i$-th bit in that encoding. For example, $(\mathit{bit}_1(6),\mathit{bit}_2(6),\mathit{bit}_3(6)) = (\zerobit,\onebit,\onebit)$. In the following, we write
$\Lambda$ for $\bigvee_{\gamma \in \Lambda} \gamma$ and $\neg\Lambda$ for $\bigwedge_{\gamma \in \Lambda}\neg \gamma$.
For an illustration of the following LTL formulas, we refer to Figure~\ref{fig:countingseq}. Formula $\firstsep(\bcsep{z}) = \neg\Lambda \wedge \neXt \bcsep{z}$ holds right before a first delimiter symbol. Similarly, $\lastsep(\bcsep{z}) = \bcsep{z} \wedge \neXt \neg\Lambda$ identifies all last delimiters and $\lastsep^-(\bcsep{z}) = \bcsep{z} \wedge \neXt\bigl(\Until{(\zerobit \vee \onebit)}{\lastsep(\bcsep{z})}\bigr)$ all second-last delimiters. Finally, we write
$\jump{z}{\psi}$ for $\neXt^{\mathit{bits}(z)+1} \psi$.

Now, $\mathit{Counter} = \mathit{Init} \wedge \mathit{Fin} \wedge \mathit{Inc} \wedge \mathit{Exit}$ is the conjunction of the following formulas:
\begin{itemize}
\item $\mathit{Init}$ says that the first binary number is always one:
\begin{center}
$\mathit{Init} = \bigwedge_{z \in Z}\Globally \bigl(\firstsep(\bcsep{z}) ~\to~ \neXt^2(\onebit ~\wedge ~\neXt (\Until{\zerobit}{\bcsep{z}}))\bigr)$
\end{center}

\item $\mathit{Fin}$ says that the last value represents $|z|$:
\begin{center}
$\mathit{Fin} = \bigwedge_{z \in Z}\Globally
\bigl(
\lastsep^-(\bcsep{z}) ~\to~ \bigwedge_{i = 1}^{\mathit{bits(z)}} \neXt^i\, \mathit{bit}_{i}(z)
\bigr)$
\end{center}

\item $\mathit{Inc}$ implements the increments:
\[
\mathit{Inc} = \bigwedge_{z \in Z}\Globally
\left(
\begin{array}{rl}
& \bigl(\bcsep{z} \wedge \neg \lastsep^-(\bcsep{z}) \wedge \neg \lastsep(\bcsep{z})\bigr)\\[0.6ex]
\to & \neXt\bigl(\Until{(\onebit \mathrel{\wedge} \jump{z}{\zerobit})}{(\zerobit \mathrel{\wedge} \jump{z}{\onebit} \mathrel{\wedge} \neXt\, \eqsuff{z})}\bigr)
\end{array}
\right)
\]
where
$\eqsuff{z} =  \Until{\bigl((\zerobit \mathrel{\wedge} \jump{z}{\zerobit}) \vee (\onebit \mathrel{\wedge} \jump{z}{\onebit})\bigr)}{\bcsep{z}}$ verifies that the suffixes (w.r.t.\ the current position) of the current and the following binary number coincide.

\item $\mathit{Exit}$ states that we do  not loop forever in the gadget from Figure~\ref{fig:nfa_binary_counting}:
\begin{center}
$\mathit{Exit} = \bigwedge_{z \in Z}\Globally \bigl(\firstsep(\bcsep{z}) ~\to~ \Future~ \lastsep(\bcsep{z}) \bigr)$
\end{center}
\end{itemize}

\noindent
Note that the gadget makes sure that, in between two delimiters $\bcsep{z}$, there are exactly $\mathit{bits}(z)$-many bits.

Now, we set $\phi' = \transl{\phi} \wedge \mathit{Counter}$.
Here, $\transl{\phi}$ simulates $\phi$ on all positions that do not contain propositions from $\Lambda$. It is inductively defined as follows:

\begin{center}
\begin{tabular}{l}
\normalsize $\transl{p} = p$ \hspace{1cm}
\normalsize $\transl{\neg p} = \neg p$ \hspace{1cm}
\normalsize $\transl{\regtest{\bowtie\,}{r}} = \regtest{\bowtie\,}{r}$ \hspace{1cm}
\normalsize $\transl{\dfreeze{r}{\psi}} = \dfreeze{r}{\transl{\psi}}$\\[1.5ex]

\normalsize $\transl{\psi_1 \vee \psi_2} = \transl{\psi_1} \vee \transl{\psi_2}$\hspace{1cm}
\normalsize $\transl{\psi_1 \wedge \psi_2} = \transl{\psi_1} \wedge \transl{\psi_2}$\hspace{1cm}
\normalsize$\transl{\neXt{\phi}} = \neXt \bigl(\Until{\Lambda}{(\neg\Lambda \wedge \transl{\phi})}\bigr)$\\[1.5ex]

\normalsize$\transl{\Until{\psi_1}{\psi_2}} = \Until{(\neg\Lambda \to \transl{\psi_1})}{(\neg\Lambda \wedge \transl{\psi_2})}$\hspace{1cm}
\normalsize$\transl{\Release{\psi_1}{\psi_2}} = \Release{(\neg\Lambda \wedge \transl{\psi_1})}{(\neg\Lambda \to \transl{\psi_2})}$
\end{tabular}
\end{center}
Note that, since $\phi$ was required to be in negation normal form, $\phi'$ is indeed in $\FFLTL$.
\end{proof}


Corollary~\ref{poca-nonempt-np} and Lemma~\ref{thm:ltl_succinct_poca_to_poca} imply the $\NEXPTIME$ upper bound for \soca model checking:

\begin{cor}\label{soca-mc-nexptime}
	\MC{\textup{\soca}}{\FFLTL} is in $\NEXPTIME$.
\end{cor}

\subsection{Lower Bounds}\label{ref:lower-blounds}

This subsection presents a matching lower bound for model checking \FFLTL. We first state $\NEXPTIME$-hardness of \MC{\textup{\ocap}}{\LTL}, which we reduce, in a second step, to \MC{\textup{\oca}}{\FFLTL}. 

\begin{thm}%
	\label{thm:ltl_succinct_POCA_completeness}
	\MC{\textup{\ocap}}{\LTL} and \MC{\textup{\socap}}{\LTL} are $\NEXPTIME$-complete.
\end{thm}
\begin{proof}
For $\NEXPTIME$-hardness of \MC{\textup{\socap}}{\LTL}, we adapt the proof of~\cite[Theorem 8]{GollerHOW10}, where G{\"o}ller et al.\ show $\NEXPTIME$-hardness of LTL model checking \ocas with parametric and succinct \emph{updates}.
In an \oca with parametric and succinct updates, the counter can be updated by either an integer (encoded in binary) or by some parameter, which, similarly to parameterized \emph{tests} in \ocap and \socap, must be assigned some concrete integer value by some parameter instantiation; note that the model in~\cite{GollerHOW10} does not allow for parameterized tests, but only for zero tests.
Since the value of the counter before an update is a priori not known, it is not possible to simulate parametric updates with parameterized tests. However, in the reduction used in the proof of~\cite{GollerHOW10}, parametric updates are preceded directly by a zero test. Hence, in this specific case, one can substitute the parametric update with a self loop increasing the counter and a parameterized test. This allows us to reuse the technique from~\cite{GollerHOW10}.

Let us give a few more details. The proof of Theorem 8 in~\cite{GollerHOW10} is a reduction from Succinct 3-SAT:\@ given a Boolean circuit $\circuit$ that encodes a Boolean formula $\psi_\circuit$ in 3-CNF, G\"oller et al.\ construct an OCA $\POCA$ with parametric and succinct updates as well as an LTL formula $\varphi$ such that there exist some parameter valuation $\gamma$ and some $\gamma$-run of $\POCA$ satisfying $\varphi$ if, and only if, $\psi_\circuit$ is satisfiable.
The \ocas $\POCA$ uses a single parameter $x$.
One can think of the value of $x$ as the encoding of a guessed variable assignment for the variables occurring in $\psi_\circuit$.
G\"oller et al.\ use $\POCA$ and $\varphi$ to verify whether this guessed assignment satisfies the Boolean formula $\psi_\circuit$.
For proving Theorem 4.7, we can almost completely rely on the proof of Theorem 8 in~\cite{GollerHOW10}. As mentioned above, all we need to change is that, in Figure 3 of~\cite{GollerHOW10}, we replace the parametric update $+\param$ by an equality test $\ptest{=}{\param}$ and add a self-loop to the start state with label $+1$.
Like in~\cite{GollerHOW10}, the value of the counter, after satisfying the parameterized test $\ptest{=}{\param}$, encodes the variable assignment of the variables occurring in $\psi_\circuit$. Satisfaction can then be verified using the same OCA gadgets and the same LTL formula as in~\cite{GollerHOW10}.
We remark that the lower bound already holds for \socap that use only one parameter $\param$ and where all parameterized tests are of the form~$\ptest{=}{\param}$.

\medskip

Membership of \MC{\textup{\ocap}}{\LTL} in $\NEXPTIME$ is by a standard argument. The given LTL formula can be translated into a B{\"u}chi automaton of exponential size. Then, we check non-emptiness of its product with the given \ocap using Corollary~\ref{cor:buechi-np}.

Moreover, \MC{\textup{\socap}}{\LTL} can be reduced to \MC{\textup{\ocap}}{\LTL} in polynomial time, using the construction from Lemma~\ref{thm:ltl_succinct_poca_to_poca}. This yields $\NEXPTIME$-hardness of \MC{\textup{\ocap}}{\LTL} and $\NEXPTIME$-membership of \MC{\textup{\socap}}{\LTL}.
\end{proof}

\begin{lem}%
	\label{lem:ocafltl-hardness}
	\MC{\textup{\oca}}{\FFLTL} and \MC{\textup{\soca}}{\FFLTL} are $\NEXPTIME$-hard.
\end{lem}

\begin{proof}
We present a reduction from \MC{\ocap}{\LTL}, which is $\NEXPTIME$-complete by Theorem~\ref{thm:ltl_succinct_POCA_completeness}.
	Let $\POCA = (Q,\params,\qinit,\Delta,\propmap)$ be an \ocap and let $\varphi$ be an LTL formula.
	We define an \oca $\POCA'=(Q',\qinit',\Delta',\propmap')$ and a sentence $\varphi' \in \FFLTL$ such that
	$\POCA \modelsex \phi$ iff $\POCA' \modelsex \phi'$.

	Without loss of generality, by~\cite{GollerHOW10} (cf.\ proof of Theorem~\ref{thm:ltl_succinct_POCA_completeness}),
	we may assume that $\POCA$ uses only one parameter $\param$ and that all parameterized tests are of the form $\mathord{=}\param$.

The idea is as follows. A new initial state $\qinit'$, in which only a fresh proposition $\mathit{freeze}$ holds, will allow $\POCA'$ to go to an arbitrary counter value, say, $v$. The \FFLTL~formula of the form $\Future\bigl(\mathit{freeze} \wedge \dfreeze{r}{\psi})$ stores $v$, which will henceforth be interpreted as parameter $x$. Hereby, formula $\psi$ makes sure that, whenever $\POCA$ performs an equality check $\mathord{=}x$, the current counter value coincides with $v$. To do so, we create two copies of the original state space: $Q \times \{0,1\}$. A transition $(q,\mathord{=}x,q')$ of $\POCA$ is then simulated, in $\POCA'$, by a $0$-labeled transition to $(q',1)$. States of the latter form are equipped with a fresh proposition $p_{\mathord{=}x}$ indicating that the current counter value has to coincide with the contents of $r$. Thus, we can set $\psi = \Globally(p_{=x} \to \mathord{=}r)$.

Formally, $\POCA'$ is given as follows. We let $Q' = (Q \times \{0,1\}) \uplus \{\qinit'\}$ and
\begin{align*}
\Delta' = & && \{(\qinit',\op,\qinit') \mid \op \in \{+1,-1\}\} ~\cup~ \{(\qinit',\zero,(\qinit,0))\}\\[1ex]
& \hspace{2mm}\cup\hspace{-4mm} && \{((q,b),0,(q,1)) \mid (q,\mathord{=}x,q') \in \Delta \text{ and } b \in \{0,1\}\}\\[1ex]
& \hspace{2mm}\cup\hspace{-4mm} && \{((q,b),\op,(q,0)) \mid (q,\op,q') \in \Delta \setminus (Q \times \{\mathord{=}x\} \times Q) \text{ and } b \in \{0,1\}\}\,.
\end{align*}
Moreover, we set $\propmap'(\qinit') = \{\mathit{freeze}\}$ as well as $\propmap'((q,0)) = \propmap(q)$ and $\propmap'((q,1)) = \propmap(q) \cup \{p_{=x}\}$ for all $q \in Q$. Finally,
$\phi' = \Future\bigl(\mathit{freeze} \wedge \dfreeze{r}{\Globally(p_{=x} \to \mathord{=}r)}\bigr) \wedge \bigl(\Until{\mathit{freeze}}{(\neg \mathit{freeze} \wedge \phi)}\bigr)$.
Note that the subformula $\Until{\mathit{freeze}}{(\neg \mathit{freeze} \wedge \phi)}$ makes sure that $\phi$ is satisfied as soon as we enter the original initial state $\qinit$ (actually, $(\qinit,0)$).

\medskip\noindent
We now prove the correctness of the above construction.

\newcommand{\OCAprime}{\POCA'}
\begin{description}\itemsep=1ex
    \item[{\bf Completeness}]
	Assume $\rho=(q_0,v_0)\pocatrans{\inst}(q_1,v_1)\pocatrans{\inst}\dots$ is a $\inst$-run of $\POCA$ for some parameter instantiation $\inst$ such that $\rho\models\varphi$.
	Using the definition of $\Delta'$, it is easy to construct from $\rho$ a run $\rho'$ of $\OCAprime$: replace every transition $(q_i,v_i)\pocatrans{\inst}(q_{i+1},v_{i+1})$ resulting from some transition $(q_i,\ptest{=}{\param},q_{i+1})\in\Delta$ by $((q_{i},b),v_i)\pocatrans{}((q_{i+1},1),v_{i+1})$, and
	replace every transition $(q_i,v_i)\pocatrans{\inst}(q_{i+1},v_{i+1})$ resulting from some transition $(q_i,\op_i,q_{i+1})\in \Delta\backslash (Q\times\{=\!\param\}\times Q)$ by $((q_{i},b),v_i)\pocatrans{}((q_{i+1},0),v_{i+1})$,
	where, in both cases, $b\in\{0,1\}$ is such that the result $\rho'$ is a run of $\OCAprime$.
	We obtain an initialized run $\rho''$ of $\OCA$ by putting in front of $\rho'$  a prefix run $\rho_{\param}$ of the form
	\begin{eqnarray*}
		(\qinit',0)\ocatrans \dots \ocatrans (\qinit',\inst(\param))\ocatrans (\qinit',\inst(\param)-1)\ocatrans \dots \ocatrans (\qinit',0)\ocatrans ((\qinit,0),0).  \\[-4mm]
		\underbrace{\hspace{1.8cm}}_{\inst(\param) \text{ times ``$+1$''} } \hspace{1.5cm} \underbrace{\hspace{4.6cm}}_{\inst(\param) \text{ times ``$-1$''} } \hspace{3.5cm}
	\end{eqnarray*}
That $\rho''\models\varphi'$ can be easily seen.
First of all,
the subformula $\dfreeze{r}{\Globally(p_{=\param} \to \mathord{=}r)}$ is satisfied at the configuration $(\qinit',\inst(\param))$: the formula stores the value $\inst(\param)$ into the register. From that position on, every time a state $(q,1)$ (with proposition $p_{=\param}$) is seen along the run $\rho''$, we know that $\OCAprime$ has just simulated a parameterized test $\ptest{=}{\param}$ of $\POCA$, hence the value of the counter is indeed equal to $\inst(\param)$. Second, the subformula  $(\neg \mathit{freeze} \wedge \phi)$ is satisfied at $((\qinit,0),0)$, by assumption and by the definition of $\propmap'$.

    \item[{\bf Soundness}]
	Assume $\rho'=(q'_0,v'_0)\ocatrans(q'_1,v'_1)\ocatrans\dots$ is a run of $\OCAprime$ such that $\rho'\models\varphi'$.
	Recall that $\propmap'(q)=\mathit{freeze}$ implies $q=\qinit'$.
	By $\rho'\models \Until{\mathit{freeze}}{(\neg \mathit{freeze} \wedge \phi)}$ we can conclude that there exists some position $j$ in $\rho'$ such that $q'_i = \qinit'$ for all $0\leq i<j$, and $q'_j=(\qinit,0)$, and
	\begin{eqnarray}
		\label{jsatisfiesvarphi}
		\rho',j\models \varphi.
		\end{eqnarray}
	By $\rho'\models \Future\bigl(\mathit{freeze} \wedge \dfreeze{r}{\Globally(p_{=x} \to \mathord{=}r)}\bigr)$ we further conclude that there exists some position $k<j$  such that \begin{eqnarray}
		\label{pequalsxatk}
		\rho',k\models \mathit{freeze} \wedge \dfreeze{r}{\Globally(p_{=x} \to \mathord{=}r)}.\end{eqnarray}
	Note that this stores the counter value $v'_k$ into the register $r$. Intuitively, $v'_k$ will correspond to the value of the parameter $x$.  We thus define $\inst(x)=v'_k$.
	By construction,
	the subrun $\rho'_{\mathsf{suffix}}$ of $\rho'$ starting in position $j$ is of the form
	\[((q_0,b_0),v_0)\ocatrans((q_1,b_1),v_1)\ocatrans((q_2,b_2),v_2)\ocatrans\dots\]
	Recall that whenever $p_{=x} \in \propmap'(s)$, then $s=(q,1)$ for some $q\in Q$.
	By (\ref{pequalsxatk}),
	$v_i=v'_k$ for all $i$ with $b_i=1$.
	Using this, it is easy to construct from $\rho'_{\mathsf{suffix}}$ a $\inst$-run $\rho$ of $\POCA$. By (\ref{jsatisfiesvarphi}), we obtain $\rho\models\varphi$.
    \qedhere
\end{description}
\end{proof}

\noindent
We conclude stating the exact complexity of model checking \oca and \soca against \FFLTL: 

\begin{thm}%
	\label{thm:main}
\MC{\textup{\oca}}{\FFLTL} and \MC{\textup{\soca}}{\FFLTL} are $\NEXPTIME$-complete.
\end{thm}

\section{Universal Model Checking}%
\label{sec:soca-universal-mc}

In this section,
we consider the \emph{universal} version of the model-checking problem (``Do all runs satisfy the formula?'').
We write $\POCA \models_\forall \phi$ if, for \emph{all} parameter instantiations $\inst$ and infinite initialized $\inst$-runs $\rho$ of $\POCA$, we have $\rho \models \phi$.
With this, universal model checking for a class $\Class$ of \socapc and a logic $\Logic \subseteq \FLTL$ is defined as follows:
\begin{center}
\begin{decproblem}
  \problemtitle{\UMC{\Class}{\Logic}}
  \probleminput{$\POCA=(Q,\params,\qinit,\Delta,\propmap) \in \Class$ and a sentence $\phi \in \Logic$}
  \problemquestion{Do we have $\POCA \models_\forall \phi$\,?}
\end{decproblem}
\end{center}
We prove in the following that, contrasting the $\nexptime$-completeness of the problem \MC{\textup{\soca}}{\FFLTL}, we have undecidability for
\UMC{\textup{\oca}}{\FFLTL}.

The proof goes a little detour over the co-flat fragment of $\FLTL$.
An $\FLTL$ formula $\varphi$ is \emph{co-flat} if its negation $\neg\varphi$ is flat. We use $\coFFLTL$ to denote the set of all co-flat $\FLTL$ formulas.

\begin{thm}
\MC{\textup{\oca}}{\coFFLTL} is undecidable.
	\end{thm}
	\begin{proof}
		A close inspection of the proof
	of~\cite[Theorem 17]{DemriLS10} (proving that \MC{\textup{\oca}}{\FLTL} is undecidable)
	reveals
 that all formulas defined are in $\coFFLTL$ so that the same proof yields undecidability of \MC{\textup{\oca}}{\coFFLTL}.
\end{proof}

A trivial reduction (simply negate the input formula) from \MC{\textup{\oca}}{\coFFLTL} proves:

\begin{thm}
\UMC{\textup{\oca}}{\FFLTL} is undecidable.
\end{thm}

\section{Parametric Timed Automata}%
\label{sec:pta}
\newcommand{\pta}{\mathcal{T}}
\newcommand{\edges}{\Delta}
\newcommand{\ptaconfigs}{\Gamma}
\newcommand{\clockval}{\nu}
\newcommand{\ptatrans}{\longrightarrow}
\newcommand{\tguard}{\pi}
\newcommand{\tTests}{\Pi}

Although \ocaps have mainly been used for deciding a model-checking problem for \ocas, they constitute a versatile tool as well as a natural stand-alone model. In fact, parameterized extensions of a variety of system models have been explored recently, among them \ocas with parameterized \emph{updates}~\cite{HaaseKOW09,GollerHOW12}, and parametric timed automata~\cite{AlurHV93,BundalaO17}, where clocks can be compared with parameters (similarly to parameterized tests in \ocaps).

In this section, we present an application of our results to the reachability problem for parametric timed automata~\cite{AlurHV93}. We refer to~\cite{Andre19} for an overview of recent advances in this field. We obtain that reachability in parametric timed automata with one parametric clock (1PTA) and closed guards (which corresponds to the discrete-time case) is in $\NEXPTIME$, matching the known lower bound. Note that this result is subsumed by~\cite{BenesBLS15} where a $\NEXPTIME$ upper bound is shown over both discrete time and continuous time, using a different proof technique.

Let $\clocks$ be a finite set of \emph{clock variables} (clocks, for short) ranging over the non-negative real numbers,  and let $\params$ be a finite set of parameters ranging over the non-negative integers.
We define \emph{tests} $\tguard$ over $\clocks$ and $\params$ by the following grammar
\begin{align*}
\tguard & ~~ ::= ~~ \clock\bowtie\param ~~\mid~~ \clock\bowtie d ~~\mid~~ \tguard\wedge\tguard
\end{align*}
where $\clock \in \clocks$, $\param\in\params$, $d\in\N$, and $\mathord{\bowtie} \in \{<,\leq,=,\geq, >\}$. We use $\tTests(\clocks,\params)$ to denote the set of all tests over $\clocks$ and $\params$.
We say  that $\tguard\in\tTests(\clocks,\params)$ is \emph{closed} if $\mathord{\bowtie}\in\{\leq,=,\geq\}$ for all atomic tests $\clock\bowtie \param$ and $\clock\bowtie d$ in $\tguard$.

\begin{defi}
A \emph{parametric timed automaton} (PTA) is a tuple $\pta = (\locs, \clocks, \params, \loc_{\mathsf{in}},\edges)$, where
\begin{itemize}
\item $\locs$ is a finite set of \emph{states},
\item $\clocks$ is a finite set of \emph{clock variables},
\item $\params$ is a finite set of \emph{parameters},
\item $\loc_{\mathsf{in}}\in \locs$ is the \emph{initial state}, and
\item $\edges \subseteq \locs \times \tTests(\clocks,\params) \times 2^\clocks \times \locs$ is the finite set of \emph{transitions}.
\end{itemize}
\end{defi}

\noindent
We say that  a clock $\clock \in \clocks$ is \emph{parametric} if there exists some transition in $\edges$ labeled with a test $\clock\bowtie\param$ that compares $\clock$ with a parameter $\param\in\params$; otherwise we say that $\clock$ is \emph{non-parametric}.
We say that $\pta$ is \emph{closed} if all tests occurring in transitions of $\pta$ are closed.

\medskip

Let $\ptaconfigs_\pta = (\locs\times{(\RP)}^\clocks)$ be the set of configurations of $\pta$, where $\RP$ is the set of non-negative real numbers.
In a configuration $(\loc,\clockval)$, the first component $\loc$ is the current state, and $\clockval$ is the current clock valuation, \ie, a mapping $\clocks\to\RP$ assigning some non-negative real number to each clock.
Like for \oca,
the semantics of a PTA $\pta$ is given with respect to a parameter instantation $\inst:\params\to\N$ in terms of a global transition relation ${\ptatrans_\inst} \subseteq\ptaconfigs_\pta\times\ptaconfigs_\pta$.

For two configurations $(\loc,\clockval)$ and $(\loc',\clockval')$, we have $(\loc,\clockval)\ptatrans_\inst (\loc',\clockval')$ if there exist a time delay $\delta\in\RP$ and some transition $(\loc,\tguard,\lambda,\loc')\in\edges$ such that
$\clockval+ \delta \models_\inst\tguard$ and $\clockval' = (\clockval+\delta)[\lambda\defeq 0]$.
Here,
\begin{itemize}
\item $\clockval+\delta$ is the clock valuation $\clockval_\delta$ defined by $\clockval_\delta(\clock) \defeq \clockval(\clock)+\delta$ for every $\clock\in\clocks$,
\item $(\clockval+\delta)[\lambda\defeq 0]$ is the clock valuation $\clockval_\lambda$ defined by $\clockval_\lambda(\clock) \defeq 0$ if $\clock\in\lambda$, and $\clockval_\lambda(\clock)\defeq(\clockval+\delta)(\clock)$, otherwise,
and
\item $(\clockval+\delta)\models_\inst\tguard$ if $(\clockval+\delta)(\clock)\bowtie \inst(\param)$ for every atomic test $\clock\bowtie\param$ in $\tguard$, and
$(\clockval+\tguard)(\clock)\bowtie d$ (for $d\in\N$) for every atomic test $\clock\bowtie d$ in $\tguard$.
\end{itemize}

\noindent
A $\inst$-run $\rho$ of $\pta$ is a finite sequence of the form $(\loc_0,\clockval_0) \ptatrans_\inst (\loc_1,\clockval_1) \ptatrans_\inst \dots \ptatrans_\inst(\loc_k,\clockval_k)$, with $k \ge 0$. We say that $\rho$ is initialized if $\loc_0 = \loc_{\mathsf{in}}$ and $\nu_0 \in {\{0\}}^\clocks$.

\medskip

The reachability problem for PTA is defined analogously to that for \oca: given some parameter instantiation $\inst$, we say that a state $\loc_f\in\locs$ is \emph{$\inst$-reachable} if there exists an initialized $\inst$-run $(\loc_0,\clockval_0) \ptatrans_\inst (\loc_1,\clockval_1) \ptatrans_\inst \dots \ptatrans_\inst(\loc_k,\clockval_k)$ such that $\loc_k=\loc_f$. We say that $\loc_f$ is \emph{reachable}, written $\Reachpred{\pta}{\loc_f}$, if it is $\inst$-reachable for some $\inst$.
\begin{center}
\begin{decproblem}
  \problemtitle{\Reach{\textup{PTA}}}
  \probleminput{A PTA $\pta = (\locs,\clocks,\params,\loc_{\mathsf{in}},\edges)$ and $\loc_f \in\locs$}
  \problemquestion{Do we have $\Reachpred{\pta}{\loc_f}$\,?}
\end{decproblem}
\end{center}
When introducing PTA in 1993~\cite{AlurHV93}, Alur, Henzinger, and Vardi proved that the problem \Reach{\textup{PTA}} is undecidable in general; however, they gave a decision procedure for PTA that use at most one parametric clock (and unboundedly many non-parametric clocks).
The computational complexity of this decision procedure is non-elementary~\cite{BundalaO17}.
When the PTA uses only a single parametric clock and no non-parametric clock, it is known that the problem is $\NP$-complete~\cite{DBLP:conf/hybrid/Miller00}. Moreover,
Bundala and Ouaknine proved that the reachability problem for \emph{closed} PTA that use at most one parametric clock (and unboundedly many non-parametric clocks) can be reduced, in exponential time, to the reachability problem for \ocapc:

\begin{propC}[\cite{BundalaO17}]
Given a closed PTA $\pta$ with one parametric clock and a state $\loc_f$ of $\pta$, one can build, in exponential time, an \ocapc $\POCA$ with a state  $q_f$ such that $\Reachpred{\pta}{\loc_f}$ iff $\Reachpred{\POCA}{q_f}$ and $|\POCA|=\mathcal{O}(2^{|\pta|})$.
\end{propC}

Based on this reduction, Bundala and Ouaknine improved the non-elementary decision procedure in~\cite{AlurHV93} for PTA with a single parametric clock  to $\tNEXPTIME$~\cite{BundalaO17} (although only for closed PTA). A $\NEXPTIME$ upper bound was then obtained by Benes et al.\, even in the continuous-time case~\cite{BenesBLS15}.

Using Theorem~\ref{thm:creachNP}, together with the hardness result appearing in~\cite{BundalaO17}, we obtain the following subcase:
\begin{thm}
\Reach{\textup{PTA}} for closed PTA with a single parametric clock is $\nexptime$-complete.
\end{thm}

\section{Conclusion}\label{sec:conclusion}

In this paper, we established the precise complexity of model checking \oca and \soca against flat \freezeLTL. To do so, we established a tight link between \ocaps and alternating two-way automata over words. Exploiting alternation further, it is likely that we can obtain positive results for one-counter games with parameterized tests. Another interesting issue for future work concerns model checking \ocaps, which, in this paper, is defined as the following question: Are there a parameter instantiation $\inst$ and a $\inst$-run that satisfies the given formula? In fact, it also makes perfect sense to ask whether there is such a run for \emph{all} parameter instantiations, in particular when requiring that all system runs satisfy a given property. It would also be worthwhile to characterize all the parameter valuations that allow one to reach a given target state. 

\bibliographystyle{alpha}
\bibliography{lit}

\appendix

\newcommand{\tree}{\chi}
\newcommand{\enc}{\mathsf{enc}}
\newcommand{\pos}{\mathsf{pos}}
\newcommand{\edge}{\Longrightarrow}
\newcommand{\edgep}[1]{\stackrel{#1}{\Longrightarrow}}

\section{Proof Details for Lemma~\ref{lemma_translation_poca_twoway}}

We have to show that $L(\twoafa) = \{w \in \paramwords{\params} \mid q_f \textup{ is } \inst_w\textup{-reachable}\}$.

\begin{description}
    \item[$\supseteq$]
Consider $w=a_0a_1a_2 \cdots \in \paramwords{\params}$ such that $q_f \textup{ is } \inst\textup{-reachable}$ where $\inst = \inst_w$.
Then, there is an initialized run $\rho = (q_0,v_0) \pocatrans{\inst} \cdots \pocatrans{\inst} (q_n,v_n)$ of $\POCA$ with $q_n = q_f$. Suppose that the global transitions are witnessed by transitions $(q_i,\op_i,q_{i+1}) \in \Delta$, for $i \in \{0,\ldots,n-1\}$. We construct a (tree) run $\tree$ of $\twoafa$ on $w$. For $v \in \N$, let $\enc(v)$ denote the unique position $i \in \N$ of $w$ such that $a_i = \adel$ and $v = |a_0 \ldots a_{i-1}|_\adel$. Thus, $\enc(v)$ is the position in $w$ that corresponds to counter value $v$.
Moreover, for $\param \in \params$, let $\pos(\param)$ be the unique position $i$ such that $a_i = \param$.
Thus, we have $\enc(\inst(x)) < \pos(x) < \enc(\inst(x)+1)$.

Starting from the root of $\tree$, which is labeled with $(s_\mathsf{in},0)$, we will first generate a finite branch that takes care of simulating $\rho$. More precisely, we create a linear sequence of nodes by replacing a global transition $(q_i,v_i) \pocatrans{\inst} (q_{i+1},v_{i+1})$, with $i \in \{0,\ldots,n-1\}$, by a sequence of configurations (proof obligations) of $\twoafa$ as follows (here, $\edge$ denotes an edge of $\tree$, possibly with the letter read at the corresponding transition):
\begin{itemize}
\item If $\op_i \in \{0\} \cup \Tests(\Params)$:
\begin{center}
$(q_i,\enc(v_i)) \edge (q_{i+1},\enc(v_{i+1}))$
\end{center}

\item If $\op_i = +1$:
\begin{align*}
(q_i,\enc(v_i))
& \edgep{\adel} (\gotoright{q_{i+1}},\enc(v_i)+1)\\
& \edgep{x_1} (\gotoright{q_{i+1}},\enc(v_i)+2)\\
& \edgep{x_2} \cdots\\
& \edgep{x_n} (\gotoright{q_{i+1}},\enc(v_{i+1}))\\
& \edgep{\adel} (q_{i+1},\enc(v_{i+1}))
\end{align*}
where $x_i\neq \adel$ are parameters.

\item If $\op_i = -1$:
\begin{align*}
(q_i,\enc(v_i))
& \edge (\gotoleft{q_{i+1}},\enc(v_i)-1)\\
& \edge \cdots\\
& \edge (\gotoleft{q_{i+1}},\enc(v_{i+1})+1)\\
& \edge (q_{i+1},\enc(v_{i+1}))
\end{align*}
\end{itemize}
Note that the resulting finite branch is both well-defined (as $w$ is a parameter word) and accepting (due to the transition $\smash{\afatrans{q_n}{\adel}{\btrue}}$ of $\twoafa$). Now, for all $i \in \{0,\ldots,n-1\}$ such that $\op_i \in \Tests(\Params) \setminus \{\zero\}$, we append another branch rooted at the node that arises from position $i$ and is labeled with $(q_i,\enc(v_i))$:
\begin{itemize}\itemsep=0.5ex
\item If $\op_i = (\ptest{=}{x})$:
\begin{align*}
(q_i,\enc(v_i))
& \edgep{\adel} (\present{\param},\enc(v_{i})+1)\\
& \edgep{x_1} (\present{\param},\enc(v_{i})+2)\\
& \edgep{x_2} \cdots\\
& \edgep{x} (\present{\param},\pos(x))
\end{align*}
where $x_i\neq x$ are parameters.
Since $\rho$ is a run, we have $v_i = \inst(x)$ and, therefore, $\enc(v_{i}) < \pos(\param) < \enc(v_{i}+1)$. Thus, this finite branch is well-defined and corresponds to a sequence of transitions of $\twoafa$. It is accepting, as $a_{\pos(x)} = \adel$ and there is a transition $\afatrans{\present{\param}}{x}{\btrue}$.

\item If $\op_i = (\ptest{<}{x})$:
\begin{align*}
(q_i,\enc(v_i))
& \edgep{\adel} (\searchn{\param},\enc(v_{i})+1) \edge \cdots \edge (\searchn{\param},\enc(v_{i}+1))\\
& \edgep{\adel} (\search{\param},\enc(v_{i}+1)+1) \edge \cdots \edge (\search{\param},\pos(x))\\
& \edgep{x} (\seen{\param},\pos(x)+1) \edge (\seen{\param},\pos(x)+2) \edge \cdots
\end{align*}
Since $\rho$ is a run, $v_i < \inst(x)$. Therefore, $\enc(v_{i}+1) < \pos(\param)$.
Thus, the branch corresponds to the transition relation of $\twoafa$.
It is infinite, since $w$ is a parameter word, and accepting, as $\seen{\param} \in S_f$ (i.e., $\seen{\param}$ is an accepting state).

\item The case $\op_i = (\ptest{>}{x})$ is similar.
\end{itemize}

\noindent
Finally, we add the branches that check that the input word is a parameter word. For every $x \in \params$, there is a branch $(s_\mathsf{in},0) \edgep{\adel} (\search{\param},0) \edge (\search{\param},1) \edge \cdots \edge (\search{\param},\pos(x)) \edgep{x} (\seen{\param},\pos(x)+1) \edge (\seen{\param},\pos(x)+2) \edge \cdots$
starting at the root of $\tree$. This infinite branch is well-defined, as $w$ is indeed a parameter word. Again, it is accepting since $\seen{\param} \in S_f$.

This concludes the construction of the accepting run $\tree$ of $\twoafa$ on $w$.

\medskip

    \item[$\subseteq$]
For the converse direction, suppose that $w=a_0a_1a_2 \cdots \in L(\twoafa)$.
Thus, there is an accepting run $\tree$ of $\twoafa$ on $w$ whose root is labeled with $(s_\mathsf{in},0)$.
We assume that $\tree$ is minimal in the sense that removing a subtree is no longer a run on $w$.
The initial transition $\smash{\afatrans{s_{\mathsf{in}}}{\adel}{(\qinit,0)} \wedge \bigwedge_{\param \in \params} (\search{\param},+1)}$ checks that the first letter is $\adel$. It also spawns a copy for each parameter $x$, checking that letter $x$ occurs exactly once.
Thus, we can assume that $w \in \paramwords{\params}$ (i.e., $w$ is a parameter word)
so that we may use the definitions $\enc(v)$ and $\pos(x)$ from the other direction.
Moreover, the root has a successor $u$ labeled with $(\qinit,0)$. Browsing through $\tree$, we successively build a $\inst_w$-run $\rho$. We start at $u$ and initialize $\rho$ to $(\qinit,0)$.

Suppose that we already constructed the run $\rho = (\qinit,0) \pocatrans{\inst_w} \cdots \pocatrans{\inst_w} (q,v)$ up to some $(q,i)$-labeled node $u$. As an invariant, we maintain that $i = \enc(v)$. If $u$ does not have any successor, then $q = q_f$, because the only transition that allows us to accept is $\smash{\afatrans{q_f}{\adel}{\btrue}}$. In that case, $q_f$ is $\inst_w$-reachable, witnessed by $\rho$. Now, suppose that $u$ has some successors.
By the definition of $\delta$, there is a transition $(q,\op,q') \in \Delta$ of $\POCA$ such that one of the following hold:
\begin{itemize}\itemsep=0.5ex
\item We have $\op = 0$ and there is a successor $u'$ of $u$ labeled $(q',i)$. Obviously, we have $(q,v) \pocatrans{\inst_w} (q',v)$ and we set $\rho' := \rho \pocatrans{\inst_w} (q',v)$.

\item We have $\op = +1$ and, starting from $u$, there is a path of the form $(q,i) \mathrel{\smash{\edgep{\adel}}} (\gotoright{q'},i+1) \edge \ldots \edge (\gotoright{q'},\enc(v+1)) \mathrel{\smash{\edgep{\adel}}} (q',\enc(v+1))$. Let $u'$ be the last node of that path. Clearly, it holds $(q,v) \pocatrans{\inst_w} (q',v+1)$, and we set $\rho' := \rho \pocatrans{\inst_w} (q',v+1)$.

\item The case $\op = -1$ is similar.

\item We have $\op = \zero$ and there is a successor $u'$ of $u$ labeled $(q',0)$ (due to $\afatrans{q}{\first}{(q',0)}$). Then, $v = 0$ and $(q,v) \pocatrans{\inst_w} (q',v)$ so that we can set $\rho' := \rho \pocatrans{\inst_w} (q',v)$.

\item We have $\op = (\ptest{=}{x})$ and, starting from $u$, there are a successor labeled with $(q',v)$ and a finite branch of the following form:
\begin{center}
$(q,i) = (q,\enc(v)) \edge (\present{\param},\enc(v)+1) \edge \cdots \edge (\present{\param},\pos(x))$
\end{center}
By the latter, we have $\enc(v) < \pos(x) < \enc(v+1)$, which implies $v = \inst_w(x)$. Thus, $(q,v) \pocatrans{\inst_w} (q',v)$ and we set $\rho' := \rho \pocatrans{\inst_w} (q',v)$.

\item We have $\op = \ptest{<}{x}$ and, starting from $u$, there are a successor $u'$ labeled with $(q',v)$ and another (accepting) infinite branch of the following form:
\begin{align*}
(q,i) = (q,\enc(v))
& \edgep{\adel} (\searchn{\param},\enc(v)+1) \edge \cdots \edge (\searchn{\param},\enc(v+1))\\
& \edgep{\adel} (\search{\param},\enc(v+1)+1) \edge \cdots \edge (\search{\param},\pos(x))\\
& \edgep{x} (\seen{\param},\pos(x)+1) \edge (\seen{\param},\pos(x)+2) \edge \cdots
\end{align*}
By the latter, we have $\enc(v+1) < \pos(x)$, which implies $v < \inst_w (x)$. Thus, $(q,v) \pocatrans{\inst_w} (q',v)$ and we set $\rho' := \rho \pocatrans{\inst_w} (q',v)$.

\item The case $\op = \ptest{>}{x}$ is similar.
\end{itemize}
We constructed a longer run $\rho'$ up to node $u'$ and continue this procedure. Since $\tree$ is accepting, we eventually find a $\inst_w$-run of $\POCA$ ending in a configuration $(q,i)$ with $q = q_f$. Thus, $q_f$ is $\inst_w$-reachable.
\end{description}

\section{Proof Details for Theorem~\ref{thm:reachNP}}
We give more detailed arguments why condition (4) in the proof of Theorem~\ref{thm:reachNP} can be  verified in polynomial time.

\smallskip

Recall from (4) that, given two configurations $(q,v)$ and $(q',v')$ with $v,v'\in D$, we need to verify whether there exists a run $(q,v)  \pocatrans{\inst}^* (q',v')$ such that, (strictly) between the configurations $(q,v)$ and $(q',v')$, the counter never takes a value in $D$.
	Assume $v=d_i$ for some $i\in\{0,\dots,n, n+1\}$. We have to distinguish the cases
	$v' = d_i$, $v' = d_{i-1}$ (if $i \ge 1$), and $v' = d_{i+1}$ (if $i \le n$). Moreover, if $v' = d_i = v$, the run can proceed below or above $v$.

Towards the case $v' = d_{i+1}$, we transform $\POCA$ into an \oca $\OCA$ as follows:
\begin{itemize}
\item remove all transitions where the label $\mathsf{op}$ is of the form
\begin{itemize}
\item $\zero$,
\item $\ptest{=}{\param}$ for some $\param\in\params$,
\item $\ptest{>}{\param}_j$ for some $j \in \{i+1,\ldots,n\}$, or
\item $\ptest{<}{\param}_j$ for some $j \in \{1,\ldots,i\}$,
\end{itemize}
\item replace $\mathsf{op}$ in all transitions by $\true$, if $\mathsf{op}$ is of the form
\begin{itemize}
\item $\ptest{<}{\param}_j$  for some $j \in \{i+1,\ldots,n\}$, or
\item $\ptest{>}{\param}_j$ for some $j \in \{1,\ldots,i\}$,
\end{itemize}
\item keep all other transitions.
\end{itemize}
It is easy to see that we have $(q,v)  \pocatrans{\inst}^* (q',v')$ in $\POCA$ such that (strictly) between the configurations $(q,v)$ and $(q',v')$ the counter never takes a value in  $D$ iff there is a $(v,v')$-run from $(q,v)$ to $(q',v')$ in $\OCA$. The latter condition can be tested in polynomial time according to Lemma~\ref{thm:galil}.

\smallskip

The other cases are handled similarly.

\section{Proof Details for Lemma~\ref{lem:buechi2reach}}
\paragraph{Soundness.}

	Assume there exist some parameter instantiation $\inst'$ and some finite $\inst'$-run $\rho'$
	in $\POCA'$ ending in $\news$.
	We distinguish the only two possible cases:

\begin{enumerate}[leftmargin=8mm,label={(\arabic*)}]\itemsep=1ex
\item The suffix of $\rho'$ is of the form
	\[(\newr,v)\pocatrans{\inst'}(\newr_1,v)\pocatrans{\inst'}(\newr_2,v)\pocatrans{\inst'}\dots\pocatrans{\inst'} (\newr_{n},v)\pocatrans{\inst'} (\news,v)\]
	for some $\newr\in \newR$ and $v\in\N$.
	Recall from the definition of $\Delta$ that the corresponding sequence of transitions is
	$(\newr,0,\newr_1)$, $(\newr_i,\ptest{>}{\param_i},\newr_{i+1})$ for all $1\leq i<n$, and $(\newr_n,\ptest{>}{x_n},\news)$. This implies $v>\inst'(\param)$ for every $\param\in\params$.
	By definition of $\newR$,
	there exists some infinite run $\hat{\rho}$ in $\POCA$
	starting in $(\newr,0)$ and visiting $q_f$ infinitely often.
	Further, $\hat{\rho}$ does not contain any transition originating from a transition in $\POCA$ with an operation of the form $\zero$, $\ptest{=}{\param}$ or $\ptest{<}{\param}$ for some $\param\in\params$.
	Hence, using the parameter instantiation defined by $\inst(\param)=\inst'(\param)$ for every $\param\in\params$,
	we can use the corresponding sequence of transitions of $\hat{\rho}$ to construct an infinite $\inst$-run of $\POCA$ that starts in $(\newr,v)$ and visits $q_f$ infinitely often:
	note that even tests of the form $\ptest{>}{\param}$ (that were replaced in $\hat{\rho}$ by $0$) are satisfied because $v>\inst(\param)$ for all $\param\in\params$.
	Concatenating this run with the prefix of $\rho'$ (until $(\newr,v)$) yields the infinite $\inst$-run we aim to construct.

\item The last transition in $\rho'$ is of the form $(\hat{q}_f,v)\pocatrans{\inst'}(\news,v)$, for some $v\in\N$, resulting from the transition $(\hat{q}_f,\ptest{=}\newxp,\news)\in \Delta'$.
	By definition of $\POCA'$,
	$\rho'$ consists of a prefix run of the form
	\[
	(\qinit,0) \pocatrans{\inst'}^* (q_f,v) \pocatrans{\inst'} (\newp,v), \]
	where the last transition results from the transition  $(q_f,\ptest{=}\newxp,\newp)\in\Delta'$, which implies $\inst'(\newxp)=v$,  and
	some suffix run of the form
	\[
		(\newp,v) \pocatrans{\inst'} (\hat{q}_1,v_1) \pocatrans{\inst'} \cdots \pocatrans{\inst'} (\hat{q}_k,v_k) \pocatrans{\inst'} (\hat{q}_f,v')\pocatrans{\inst'}(\news,v')
	\]
	where the last transition again results from the transition  $(\hat{q}_f,\ptest{=}\newxp,s)\in\Delta$, which implies $v'=v$.
	Define $\inst$ by $\inst(\param)=\inst'(\param)$ for every $\param\in\params$.
	The construction of $\POCA'$ implies that in $\POCA$  there must be finite $\inst$-runs
	$\rho_{\mathsf{prefix}}$ and $\rho_{\mathsf{suffix}}$ of the form
	\[
		\rho_{\mathsf{prefix}}= (\qinit,0) \pocatrans{\inst}^* (q_f,v)
	\]
	and
	\[
		\rho_{\mathsf{suffix}} = (q_f,v) \pocatrans{\inst} (q_1,v_1) \pocatrans{\inst} \cdots \pocatrans{\inst} (q_k,v_k) \pocatrans{\inst} (q_f,v),
	\]
	yielding an infinite $\inst$-run defined by $\rho_{\mathsf{prefix}} {(\rho_{\mathsf{suffix}})}^\omega$, visiting $q_f$ infinitely often.
\end{enumerate}

\paragraph{Completeness.}

	Assume there exist some parameter instantiation $\inst$ and some infinite $\inst$-run $\rho$ in $\POCA$ visiting $q_f$ infinitely often.
	We distinguish two (not necessarily disjoint) cases:

\begin{enumerate}[leftmargin=8mm,label={(\arabic*)}]\itemsep=1ex
\item Run $\rho$ visits $q_f$ infinitely often with the same counter value $v$, for some $v\in\N$.
	Define $\inst'$ by $\inst'(\param)=\inst(\param)$ for all $\param\in\params$, and $\inst'(\newxp)=v$.
	It can be easily seen that, from $\rho$, we can construct a finite $\inst'$-run $\rho'$ that ends in $\news$:
	when, in $\rho$, the configuration $(q_f,v)$ is visited for the first time, we use the transition $(q_f,\ptest{=}\newxp,p)$, from which $\POCA'$ can simulate $\rho$ until it meets $(\hat{q}_f,v)$ for the next time.

\item Run $\rho$ visits $q_f$ infinitely often with unbounded counter values.
	Define $M=\max\{\inst(\param)\mid\param\in\params\}$ to be the maximum value that a parameter is instantiated with.
	We distinguish two cases:
	\begin{enumerate}[label={(\alph*)}]\itemsep=0.5ex
\item 	Run $\rho$ visits infinitely often a counter value $m\le M$.
	Then, $\rho$ must necessarily contain a subrun of the form
	\[ (q,m) \pocatrans{\inst}^* (q_f,v) \pocatrans{\inst}^* (q,m)\] for some $q\in Q$ and $v \in \N$.
	Note that, then, there must necessarily exist an infinite $\inst$-run that visits $(q_f,v)$ infinitely often, so that we can proceed with case (1) above.

\item There exists some configuration $(\newr,v)$ in $\rho$, where $v>M$, such that the infinite $\inst$-run $\rho'$ starting in $(\newr,v)$ never visits the counter value $M$ anymore. Hence, $\rho'$ cannot contain any transitions with operations of the form $\zero$, $\ptest{=}{\param}$, or $\ptest{<}{\param}$ for some $\param\in\params$,
	and transition guards of the form $\ptest{>}{\param}$ for some $\param\in\params$ are satisfied. Clearly, $\rho'$ is visiting $q_f$ infinitely often, and hence $\newr\in \newR$.
	It is easily seen that, from $\rho$, we can construct a finite $\inst'$-run (where the fresh parameter $\newxp$ is indeed never used) that ends in $\news$ using the sequence of transitions $(\newr,0,r_1), (\newr_1,\ptest{>}{\param}_1,r_2),\dots,(\newr_n,\ptest{>}{\param}_n,\news)$.
\end{enumerate}
\end{enumerate}

\end{document}